\documentclass[useAMS,usenatbib]{mn2e}
\usepackage{graphicx}
\usepackage{amssymb}
\usepackage{lscape}
\usepackage{ulem}

\def\kms{km~s$^{-1}$}
\def\cm{cm$^{-2}$}
\def\lya{Ly$\alpha$}
\def\nhi{$N$(H\,{\sc i})}
\def\hi{H\,{\sc i}}
\def\si2{Si\,{\sc ii}}
\def\fe2{Fe\,{\sc ii}}
\def\al2{Al\,{\sc ii}}
\def\zn2{Zn\,{\sc ii}}
\def\c2s{C\,{\sc ii}$^{\star}$}

\title[A search for DLAs suitable for 21~cm follow-up]{A search for 
damped Lyman-$\alpha$ systems towards radio-loud 
quasars I: The optical survey.\thanks{Based on observations made with 
(1) ESO Telescopes at the Paranal Observatories under programme
IDs 075.A-0015(B) and 077.A-0139(A,B), (2) the Gemini-North Observatory
under programme IDs GN-2005B-Q-60 and GN-2006A-Q-42 and (3) the William
Herschel Telescope under programme IDs W/06A/P3 and W/06B/P5.}}

\author[Ellison et al.] {Sara L. Ellison$^1$, Brian A. York$^1$,  
Max Pettini$^2$, Nissim Kanekar$^3$.\\
$^1$Department of Physics and Astronomy, University of Victoria, Victoria, B.C., V8P 1A1, Canada\\
$^2$Institute of Astronomy, Madingley Rd., Cambridge, CB3 0HA, UK\\
$^3$National Radio Astronomy Observatory, 1003 Lopezville Road, Socorro, NM 
    87801, USA}

\begin{document}

\maketitle

\begin{abstract}

We present the results from the optical component of a survey
for damped Lyman-$\alpha$ systems (DLAs) towards radio-loud
quasars.  Our quasar sample is drawn from the Texas radio 
survey with the following primary selection criteria: $z_{\rm em} \ge 2.4$, 
optical magnitudes $B \le 22$ and 365~MHz flux density 
$S_{365} \ge 400$ mJy.  We obtained spectra for a sample of
45 QSOs with the William Herschel Telescope, Very Large Telescope
and Gemini-North, resulting in a survey redshift path $\Delta z = 38.79$.
We detect nine DLAs and one sub-DLA, with a mean absorption redshift
$\langle z \rangle = 2.44$.  The DLA number density is $n(z) =
0.23^{+0.11}_{-0.07}$, in good agreement with the value derived
for DLAs detected in the Sloan Digital Sky Survey at this redshift. The
DLA number density of our sample
is also in good agreement with \textit{optically-complete} 
radio-selected samples, supporting previous claims that
$n(z)$ is not significantly affected by dust obscuration bias.
We present \nhi\ column density determinations and metal line equivalent width
measurements for all our DLAs.  The low frequency flux density selection
criterion used for the quasar sample implies that all absorbers will be 
suitable
for follow-up absorption spectroscopy in the redshifted \hi\ 21\,cm line.  
A following paper 
(Kanekar et al.) will present \hi\ 21\,cm absorption studies of, and spin
temperature determinations for, our DLA sample.

\end{abstract}


\section{Introduction}

Damped Lyman-$\alpha$ systems (DLAs) currently provide the
most complete means to study the gas-phase interstellar
medium (ISM) of high redshift galaxies.  Many hundreds of
DLAs have now been identified (e.g. Prochaska, Herbert-Fort
\& Wolfe 2005), with chemical abundances derived for
$\sim$ 150 from high resolution optical spectroscopy
(Prochaska et al. 2007).  Such spectroscopy also permits
studies of the gas velocity fields, although the interpretation
can be complicated (Prochaska \& Wolfe 1997; Haehnelt, Steinmetz
\& Rauch 1998;
Maller et al. 2001).  Few other physical quantities
can be derived for DLAs in a wholesale manner.  Although
a handful of DLAs have been subjected to detailed modelling of
the ionization state of the ISM (e.g. Prochaska et al. 2002; Lopez et al.
2002; Tripp et al. 2005), have galactic
counterparts identified (M\o ller et al. 2004 and references
therein) or have size estimates
(Lopez et al. 2005, but see Ellison et al. 2007),
there are relatively few tools at our disposal to
determine the physical characteristics of high redshift
DLA galaxies.  One example of a technique that relies on well-understood
physics and has been widely applied to the Milky Way and Magellanic
Clouds (e.g. Tumlinson et al. 2002), but is only recently realising 
its potential at high redshift, is the study of molecular hydrogen 
(Ge \& Bechtold 1997; Ge, Bechtold \& Kulkarni 2001;
Ledoux et al. 2003; Srianand et al 2005).  For example,
in two DLAs which exhibit molecular H$_2$ absorption spanning 
a number of rotational levels, Noterdaeme et al. (2007a,b) have
been able to compute the properties of the radiation
fields as well as gas phase temperatures and densities.  Another 
technique that can be used to probe the properties of absorption systems
has been developed by Wolfe, Prochaska \& Gawiser (2003) who
use detections of \c2s\ (absorption from the 
excited fine-structure level of the ground state)
to infer the star formation
rate (SFR) in DLA galaxies. The downsides of these techniques are 
that they are both model-dependent and rely on the detection of uncommon
species, so that they may present a somewhat biased
view of the high redshift ISM.  For example, the molecular hydrogen
fraction is greater than $10^{-4.5}$ in only 4\% of DLAs and sub-DLAs 
(19 $<$ log \nhi\ $<$ 20.3) when the metallicity is less than 1/20 of 
the solar value.  This percentage rises to 35\% for absorbers whose
metallicity is greater than 1/20 of the solar value, indicating
that molecule formation/survival is more successful in metal-rich, dustier
systems  (Noterdaeme et al. 2008).  \c2s\ is detected in about 60\%
of DLAs (Wolfe et al. 2008).  It has also been shown by Wolfe et al. 
that the distribution of cooling rates (which is directly
determined from the column densities of \hi\ and \c2s) is bimodal.  
These authors argue that this bimodality
is due to an underlying difference in the mode of star formation, 
analogous to the bimodality that leads to the blue and red sequences 
in galaxies at low redshift.  The strength of (and therefore ability to
detect) \c2s\ absorption [for a given \nhi] therefore depends on
the properties of the galaxy.

\hi~21\,cm absorption studies of DLAs towards radio-loud 
quasars provide an alternative probe of the state of the ISM 
in high redshift galaxies (see Kanekar \& Briggs 2004 for
a review), through measurements of the \hi\ spin temperature. 
The latter quantity is determined by combining measurements 
of the \hi\ column density (determined, in DLAs, from the Lyman-$\alpha$ 
absorption line) and the \hi\ 21\,cm optical depth. In typical Galactic 
clouds, the spin temperature, T$_{\rm s}$, 
of cold \hi\ clouds (where collisions 
are sufficiently frequent to thermalise the \hi\ 21\,cm transition) 
is approximately the same as the gas kinetic temperature 
(e.g. Kulkarni \& Heiles 1988); conversely, T$_{\rm s}$ is expected to be 
lower than the kinetic temperature for the warm phase of \hi\ (e.g. 
Liszt  2001). For multi-phase lines of sight, the measured
spin temperature can be shown to be the column-density-weighted
harmonic mean of the spin temperatures of different phases 
(and is thus biased towards the phase with the lowest temperature). 
For example, a sightline with 10\% of its gas at 100 K
and 90\% of its gas at 8000 K (typical temperatures of the
cold and warm neutral media, CNM and WNM respectively)
would have a measured spin temperature T$_s \sim$ 900 K.
\hi\ 21\,cm absorption studies of DLAs towards compact quasars thus 
provide information on the {\it distribution of gas in 
different temperature phases} and can be used to study the evolution 
of this temperature distribution with redshift (e.g. Wolfe \& Davis 
1979; Wolfe, Briggs \& Jauncey 1981; 
de Bruyn et al. 1996; Carilli et al. 1996; 
Lane et al. 1998;  Chengalur \& Kanekar 2000; 
Kanekar \& Chengalur 2001; Kanekar \& Chengalur 2003).

DLA spin temperatures may also yield other clues to the character
of high redshift galaxies, beyond the direct probe of the gas 
temperature.  For example, there is evidence that T$_{\rm s}$ is 
connected with galaxy size and morphology (e.g. Chengalur \& Kanekar 2000).
In the local universe, large disk galaxies like the Milky Way and 
M31 tend to have significant cold gas fractions, yielding 
low harmonic mean spin temperatures , T$_{\rm s} \lesssim 300$~K (Braun \& 
Walterbos 1992; Braun 1997). Conversely, dwarf galaxies 
appear to have far larger fractions of the warm phase of \hi\, with
T$_{\rm s} \gtrsim 1000$~K (e.g. Young \& Lo 1997). Kanekar \& Chengalur 
(2003) have argued that a similar connection between T$_{\rm s}$ and morphology
exists in DLAs, at least out to $z \sim 0.7$.  If this relationship 
persists to higher redshifts, the spin temperature would be an excellent 
probe of galaxy morphology at redshifts where direct imaging of galaxies 
is extremely difficult.  There is also tentative evidence for an anti-correlation
between metallicity and spin temperature (Kanekar \& Briggs 2004;
Kanekar et al. in preparation), further suggesting that the gas temperature distribution
in a typical DLA may be determined by global factors like mass, SFR, metallicity, etc. Finally, understanding whether the ISM of typical 
DLAs is dominated by the CNM or WNM has significant implications 
for modelling SFRs at high $z$ (e.g. Wolfe et al. 2003).

Although the measurement of DLA spin temperatures holds great promise 
as a probe of the ISM in high $z$ galaxies, there are a number of problems that
need to be addressed.  For example, although the C\,{\sc ii}$^{\star}$  model of Wolfe et al. (2003) 
requires about half of the high $z$ DLA population to have sizeable fractions of cold 
\hi, this is apparently contradicted by the high spin temperatures universally measured 
in $z > 2$ DLAs prior to our survey (e.g. Carilli et al. 1996; Kanekar \& Chengalur 2003; 
Kanekar et al 2006, 2007).  Only at $z<1$ have spin temperatures T$_s \lesssim  300$~K been obtained 
(see Table~3 of 
Kanekar \& Chengalur 2003).  The reason behind the apparent dearth of low spin 
temperatures at high redshift has been the cause of considerable debate.  One 
possibility is that the galaxy population at $z>2$ is dominated by dwarfs, 
whereas at low redshift, a more diverse mix of morphologies exists (e.g. Kanekar \& 
Chengalur  2001).  Alternatively, it has been suggested 
that the high spin temperatures seen in high-$z$ DLAs arise as an artefact of 
either low covering factors (e.g. Curran et al. 2005; Curran \& Webb 2006) 
or differences between the 
optical and radio lines of sight (Wolfe et al. 2003).

One of the problems in untangling the issues described above is the small number of \hi\ 21\,cm
absorption studies at $z>2$.  After nearly thirty years of searches prior to
the present survey, there have been only five detections of \hi\ 21\,cm absorption in 
$z \gtrsim 1.8$ DLAs (Wolfe \& Davis 1979; Wolfe et al. 1981; Wolfe et al. 1985; Briggs, Brinks \& Wolfe 1997; 
Kanekar et al. 2006, 2007), with $\sim 12$~non-detections of absorption, yielding 
strong lower limits ($\gtrsim 1000$~K) on the spin temperature (e.g. Kanekar \& 
Chengalur 2003). Most of these results have been obtained in the last 
decade, as the result of improved frequency coverage on telescopes like the 
Giant Metrewave Radio Telescope (GMRT) and the Green Bank Telescope (GBT).

The primary reason for the lack of spin temperature measurements in high-$z$ DLAs
is simply the paucity of known DLAs towards radio-loud QSOs. For example, 
only two of the $\sim 500$ DLAs detected in the SDSS-DR3 sample of Prochaska 
et al. (2005) lie towards QSOs with sufficient low-frequency flux density 
for follow-up 21\,cm absorption studies. In addition, only two DLA surveys have specifically 
targeted radio-loud quasars, the Complete Optical and Radio Absorption Line System (CORALS) 
survey by Ellison et al. (2001) and the UCSD survey of Jorgenson et al. (2006).  
However, both of these surveys selected quasars on the basis of their high-frequency ($2.7-5$~GHz)
flux density.  Unfortunately, in many cases, the DLA detections occurred towards quasars 
whose flux density at low frequencies (corresponding to the redshifted \hi\ 21\,cm line) is
insufficient for absorption follow-up. A significant improvement of T$_{\rm s}$ statistics 
clearly requires a dedicated survey for DLAs towards quasars selected at frequencies 
of 300 -- 500 MHz.

We have conducted precisely such an optical survey, specifically designed to
detect DLAs towards radio-loud quasars selected at low frequencies. In theory 
(i.e. excluding issues related to local radio frequency interference),
it should be possible to follow up every DLA detected in this survey with
a search for redshifted \hi\ 21\,cm absorption. In this paper, we describe the 
optical portion of the survey, including the survey design and sample
selection (section \ref{sample_sec}), observations and data reduction 
(sections \ref{obs_sec} and \ref{data_sec}) and the DLA detections, fits and 
statistical properties (section \ref{dla_sec}). The survey has already yielded 
the first detection of a low spin temperature DLA at $z \gtrsim 2$ (York et al. 
2007). A following paper (Kanekar et al., in preparation) will present the 
\hi\ 21\,cm observations and inferred spin temperatures for the full sample of 
DLAs detected in the survey.

\section{Sample selection}\label{sample_sec}

Our goal is to identify DLAs from optical, ground-based
spectra towards radio-loud quasars that will be suitable
for follow-up spectroscopy in the redshifted \hi\ 21\,cm line.  In order to proceed
efficiently with this search, we must therefore first identify
a sample of quasars that is well characterised at both
optical and radio wavelengths.  The quasars must obviously
be radio-loud, but we must specifically be careful to select targets
that have sufficient flux at 1420/(1+$z_{\rm abs}$) $\sim$
300 -- 500 MHz to permit later searches for redshifted \hi\ 21\,cm absorption.  
We selected the Texas
radio survey (Douglas et al. 1996) as the most suitable for our purpose,
due to its low frequency and wide area sky coverage.
In brief, the Texas (TXS) survey was conducted with the Texas
interferometer at 365 MHz between 1974 and 1983 and contains
almost 67,000 discrete sources with arcsecond positional
precision.  The
survey covers declinations between $-$35.5 degrees to +71.5
degrees and is complete to 0.4 Jy and 80\% complete to
0.25 Jy (Douglas et al. 1996).

In order to select a sample of high redshift quasars from
within the TXS survey, we used the NASA Extragalactic Database
(NED) with the following search criteria:

\begin{itemize}
\item TXS sources classified as QSOs
\item Redshifts $z_{\rm em} \ge 2.40$
\item 365 MHz flux density $S_{\rm 365} \ge 400$ mJy
\end{itemize}

This search yielded a total of 123 targets.  From within this list,
we excluded objects that we considered too faint at optical wavelengths
for efficient follow-up (typically $B > 22$) and any
targets which had already been included in previous DLA 
searches\footnote{8 of the QSOs in our survey have subsequently
also appeared in the DLA survey of Jorgenson et al. (2006).}.
In order to fill the ranges in right ascension suitable
for scheduled observing runs, we relaxed these criteria
in a few cases to include additional objects.

Observing time for this project was allocated at Gemini-North, 
the William Herschel Telescope (WHT) and the Very Large Telescope (VLT) 
in observing semesters from early 2005 until late 2006.
The total telescope time allocated to this project
amounted to 17.2 hours at the VLT, 32.2 hours on Gemini-North
(of which 28 were completed)
and 5 nights at the WHT (see the next section and
Table \ref{tel_time}).  
Targets with $B$-band magnitudes
brighter than 20.0 and declinations $\delta > -10$
were typically observed with the WHT.  Fainter targets
at northerly declinations were assigned to Gemini,
whilst all targets with $\delta < -10$ were observed
with the VLT.
With these allocations, we were able to complete
observations of 45 QSOs for our DLA search.
The final list of quasars in our survey
is given in Table \ref{optical_obs}.

\section{Observations}\label{obs_sec}

Observing procedures differed from telescope to
telescope.  At both Gemini-North and the VLT, observations
were carried out in queue/service mode, according to
the practices (e.g. scheduling and calibration plans)
of those particular observatories.  The WHT observations
were carried out in classic/visitor mode.  We briefly
describe the instrument configurations, run conditions 
and notable observing strategies for each telescope.

\begin{center}
\begin{table}
\caption{Telescope time allocations}
\begin{tabular}{lccc}
\hline
Telescope & Semester  & Allocation & Actual\\ \hline
WHT & 2005A & 3 nights & 3 nights\\
WHT & 2006B & 2 nights & 1.5 nights \\
Gemini-North & 2005B  & 19 hours & 19 hours \\
Gemini-North & 2006A  & 13.2 hours & 9 hours \\
VLT & P75 & 2.2 hours & 2.2 hours\\
VLT & P77 & 15 hours & 15 hours\\  
\hline 
\end{tabular}\label{tel_time}
\end{table}
\end{center}

\subsection{WHT observations}

\begin{center}
\begin{table*}
\caption{QSO observation log for optical observations}
\begin{tabular}{lccccccc}
\hline
QSO & $z_{\rm em}$ &  $B$ mag  & 365 MHz flux    & Telescope  & Obs. date &  Exposure &  S/N \\
    &              &           &  density (Jy)  &            &           & time (s) &  pix$^{-1}$ \\
\hline
TXS0017$-$307 &   2.67 & 20.0 &  0.496 $\pm$ 0.051  & VLT &      2006 Jul. 27 &            1000 &      8-80 \\
TXS0211$+$296 &   2.87 & 20.5 &  0.412 $\pm$ 0.021  & Gemini &   2005 Nov. 1,2 &         7200 &      7-100 \\
TXS0214$-$011 &   2.45 & 20.0 &  0.562 $\pm$ 0.045  & WHT &      2006 Nov. 23,24 &         7200 &      5-13 \\
TXS0222$+$185 &   2.69 & 20.1 &  0.425 $\pm$ 0.041  & WHT &      2006 Nov. 24 &            7200 &      4-12 \\
TXS0223$+$341 &   2.91 & 22.0 &  3.799 $\pm$ 0.034  & Gemini &   2006 Jul. 4,26,27 &      10800 &     6-40 \\
TXS0229$+$230 &   3.42 & 22.0 &  0.581 $\pm$ 0.033  & Gemini &   2006 Jul. 28 &            10800 &     6-60 \\
TXS0258$+$058 &   2.31 & 20.8 &  0.608 $\pm$ 0.028  & WHT &      2006 Nov. 24 &            3600 &      8-15 \\
TXS0304$-$316 &   2.54 & 18.9 &  0.718 $\pm$ 0.058  & VLT &      2006 Aug. 16 &            1000 &      7-65 \\
TXS0311$+$430 &   2.87 & 21.5 &  5.483 $\pm$ 0.052  & Gemini &   2005 Oct. 30, Nov. 2,3 &14400 &     8-20 \\
TXS0351$+$187 &   2.71 & 20.0 &  0.826 $\pm$ 0.021  & WHT &      2006 Nov. 24 &            7200 &      8-20 \\
TXS0441$+$106 &   2.40 & 19.6 &  0.856 $\pm$ 0.027  & WHT &      2006 Nov. 24 &            5400 &      8-22 \\
TXS0609$+$607 &   2.70 & 20.0 &  1.562 $\pm$ 0.033  & WHT &      2006 Nov. 24 &            7200 &      6-18 \\
TXS0620$+$389 &   3.46 & 20.0 &  2.233 $\pm$ 0.081  & WHT &      2006 Nov. 24 &            7200 &      7-16 \\
TXS0723$+$488 &   2.46 & 20.2 &  0.646 $\pm$ 0.019  & Gemini &   2005 Nov. 2,3 &         3600 &      6-60 \\
TXS0859$+$433 &   2.41 & 20.7 &  0.606 $\pm$ 0.052  & Gemini &   2005 Nov. 4 &            7200 &      7-110 \\
TXS0902$+$490 &   2.69 & 18.0 &  0.976 $\pm$ 0.024  & WHT &      2005 Apr. 17 &            1800 &      10-40 \\
TXS0907$+$258 &   2.74 & 18.2 &  0.453 $\pm$ 0.032  & WHT &      2005 Apr. 16,18 &         1800 &      9-50  \\
TXS0930$+$493 &   2.58 & 18.8 &  0.488 $\pm$ 0.021  & WHT &      2005 Apr. 18 &            3600 &      6-25 \\
TXS0935$+$397 &   2.49 & 20.2 &  0.445 $\pm$ 0.039  & Gemini &   2005 Nov. 29 &            3600 &      7-50 \\
TXS1013$+$524 &   2.45 & 19.9 &  0.846 $\pm$ 0.038  & WHT &      2005 Apr. 18 &            3600 &      4-13 \\
TXS1017$+$109 &   3.15 & 18.7 &  1.642 $\pm$ 0.034  & WHT &      2005 Apr. 16,17 &         3600 &      17-30 \\
TXS1025$-$264 &   2.66 & 18.4 &  1.735 $\pm$ 0.128  & VLT &      2006 Apr. 5 &            1000 &      25-75 \\
TXS1048$+$347 &   2.52 & 20.4 &  0.574 $\pm$ 0.022  & Gemini &   2005 Dec. 4,22,24 &      7200 &      13-50 \\
TXS1053$+$704 &   2.49 & 19.3 &  0.447 $\pm$ 0.023  & WHT &      2005 Apr. 17 &            5400 &      5-20 \\
TXS1214$+$348 &   2.64 & 19.4 &  0.578 $\pm$ 0.052  & WHT &      2005 Apr. 18 &            3600 &      12-40 \\
TXS1214$+$588 &   2.54 & 19.6 &  0.710 $\pm$ 0.058  & WHT &      2005 Apr. 18 &            3600 &      14-35 \\
TXS1239$+$376 &   3.81 & 20.0 &  0.469 $\pm$ 0.019  & WHT &      2005 Apr. 17,18 &         7200 &      5-53  \\
TXS1313$+$200 &   2.47 & 18.0 &  0.844 $\pm$ 0.025  & WHT &      2005 Apr. 19 &            3600 &      10-30 \\
TXS1358$+$046 &   2.55 & 21.3 &  0.771 $\pm$ 0.027  & VLT &      2006 Apr. 3 &            4000 &      10-75 \\
TXS1427$+$263 &   2.91 & 18.3 &  0.798 $\pm$ 0.029  & WHT &      2005 Apr. 17 &            2700 &      10-40 \\
TXS1442$+$101 &   3.52 & 19.0 &  2.490 $\pm$ 0.210  & WHT &      2005 Apr. 18 &            3600 &      20-66 \\
TXS1445$-$161 &   2.41 & 19.9 &  2.494 $\pm$ 0.053  & VLT &      2006 Apr. 3 &            1000 &      10-62 \\
TXS1452$+$502 &   2.84 & 19.7 &  3.439 $\pm$ 0.146  & WHT &      2005 Apr. 17,18 &         4800 &      28-45 \\
TXS1455$+$348 &   2.73 & 19.8 &  0.739 $\pm$ 0.051  & WHT &      2005 Apr. 19 &            5400 &      6-15 \\
TXS1539$+$477 &   2.80 & 18.2 &  0.645 $\pm$ 0.018  & WHT &      2005 Apr. 17 &            3600 &      5-37 \\
TXS1557$+$032 &   3.89 & 21.5 &  0.543 $\pm$ 0.070  & VLT &      2005 Apr. 10 &            4000 &      6-46  \\
TXS1701$+$379 &   2.45 & 20.0 &  0.537 $\pm$ 0.053  & WHT &      2005 Apr. 19 &            5400 &      11-45 \\
TXS1722$+$526 &   2.51 & 18.0 &  0.797 $\pm$ 0.021  & WHT &      2005 Apr. 18 &            2200 &      15-70 \\
TXS2015$+$657 &   2.84 & 21.0 &  0.884 $\pm$ 0.062  & Gemini &   2006 May 23,27 &         5400 &      9-140 \\
TXS2039$+$187 &   3.05 & 21.5 &  1.865 $\pm$ 0.027  & VLT &      2006 May 25,28 &         4000 &      7-140  \\
TXS2127$+$348 &   2.40 & 19.5 &  0.515 $\pm$ 0.037  & WHT &      2006 Nov. 24 &            5400 &      15-30 \\
TXS2131$-$045 &   4.34 & 21.5 &  0.622 $\pm$ 0.058  & VLT &      2006 May. 25 &            4000 &      16-35  \\
TXS2211$-$251 &   2.50 & 20.1 &  2.829 $\pm$ 0.132  & VLT &      2006 May 28, Jun. 21 &     1000 &      8-65 \\
TXS2320$-$312 &   2.54 & 18.8 &  1.592 $\pm$ 0.051  & VLT &      2006 May 15 &            1000 &      8-106 \\
TXS2338$+$042 &   2.59 & 19.9 &  6.067 $\pm$ 0.147  & WHT &      2006 Nov. 23 &            7200 &      2-17 \\
\hline 
\end{tabular}\label{optical_obs}
\end{table*}
\end{center}

\begin{center}
\begin{table*}
\caption{Instrument set-ups}
\begin{tabular}{lccccccc}
\hline
Telescope &     Run    & Instrument  &  Grating &   Slit width &    CCD      &  Resolution &    Wavelength  \\
          &     dates  &             &  or grism&   (arcsec)  &     binning  &  FWHM (\AA) &    coverage (\AA)   \\
\hline
WHT &           2005 Apr. 16-18 & ISIS Blue Arm & R300B & 0.7-1.0 & 1x1 &  2.5 &  3200-6200 \\
WHT &           2006 Nov. 23-24 & ISIS Blue Arm & R300B & 1.0 &     1x1 &  2.5 &  3200-6200 \\
Gemini-North &  2005B &           GMOS-N &        B600 &  1.0 &     2x2 &  4.6 &  3700-6000 \\
Gemini-North &  2006A &           GMOS-N &        B600 &  1.0 &     2x2 &  4.6 &  3700-6000 \\
VLT &           Period 75  &            FORS2 &         600B &  1.0 &     2x2 &  4.7 &  3500-6300 \\
VLT &           Period 77  &            FORS2 &         600B &  1.0 &     2x2 &  4.7 &  3500-6300 \\
\hline 
\end{tabular}\label{optical_setup}
\end{table*}
\end{center}

The Intermediate dispersion Spectrograph and Imaging 
System (ISIS) was used for our WHT observations.
ISIS is a dual arm spectrograph whose red and blue
channels can be fed through a selection of dichroics.
Our observations only made use of the blue arm, since
the entire Lyman-$\alpha$ forest could be covered with a single
setting for our targets.  We used the R300B grating
which is blazed at 4000 \AA\ and yields a total
wavelength coverage of 3539 \AA.  However, there
is significant vignetting with the $4096\times2048$
$13\,\mu$m EEV CCD detector, such that the total
unvignetted spectral coverage is approximately
2800 \AA\ per tilt.  The central wavelength for
all targets observed with ISIS was 4600 \AA,
yielding a total unvignetted
wavelength coverage of 3200 -- 6000 \AA, although
in practice a high fraction of the flux is recovered out to
6200 \AA.  In some cases, a reasonable signal
was also recovered slightly blueward of 3200 \AA.
Table \ref{optical_setup} summarises the relevant 
instrument setup information.  

The WHT observations were conducted during two runs:
3 nights in April 2005 and 2 nights in November
2006.  The April run experienced mixed conditions, with some
high cirrus, high humidity and seeing of 1.0 -- 1.2 arcsec
at the start of the run, improving to 0.5 arcsec and
clear skies by the end of the third night.  The slit
width was chosen to track the conditions and varied
from 1.0 to 0.7 arcsec.
The November run experienced poorer conditions:
seeing was 0.9 -- 1.0 arcsec throughout, so the slit
width was fixed at 1.0 arcsec.  The second night
of this run suffered from high humidity and the
latter half was lost for this reason.
Previous experience with ISIS revealed that the drift
in wavelength calibration through the night can be significant,
up to $\sim$ 2 \AA.  We therefore took exposures for
wavelength calibration (with the internal CuAr and
CuNe lamps) after each target slew.  We repeated
the arc observations after every two on-target
exposures of typically 1800 seconds each.

\subsection{Gemini-North observations}

We utilized the Gemini Multi-Object Spectrograph (GMOS)
on Gemini-North.  We used the B600 grating which is
blazed at 4610 \AA\ and yields a total (theoretical) wavelength
coverage of 2760 \AA\ per tilt.  Our central wavelength
was set to 4590 \AA, with 30 \AA\ offsets between
exposures of a given target in order to cover the
gaps between the 3 CCDs.  
Despite the theoretical coverage
at the blue end of the spectral range, the very low
efficiency of the telescope and instrument at these
wavelengths meant
that there was essentially zero flux blueward of
3700 \AA.  The effective wavelength coverage of
our spectra suitable for a DLA search was therefore 
typically 3700 -- 6000 \AA.
Table \ref{optical_setup} summarises the relevant 
instrument setup information.  

Observations were obtained during semesters 2005B
and 2006A in queue mode.  The observing conditions
were set to be conducted in seeing conditions
in the 85$^{\rm th}$ percentile (1.0 arcsec or better)
and with cloud coverage in the 70$^{\rm th}$ percentile
(0.3 magnitudes of extinction or less) with a fractional
lunar illumination (FLI) less than 50\%.  A 1.0 arcsec
slit was used throughout, which, combined with 2$\times$2
on-chip binning, resulted in a FWHM resolution of
approximately 4.6 \AA.  Exposures for wavelength calibration
were obtained with a CuAr lamp, with one exposure
taken at each central wavelength setting for each set
of exposures.  In practice, the observing sequence executed
the science exposures centred at 4590 \AA, followed by
a wavelength calibration at that wavelength.  The new
grating tilt was then applied, a wavelength calibration
taken at the central wavelength of 4620 \AA, and then
the science exposures at this wavelength.

\subsection{VLT observations}

We used the FOcal Reducer and low dispersion Spectrograph
2 (FORS2) at the VLT (UT1) with the 600B grism.  The
central wavelength of FORS2 is set by the combination
of the grism and slit.  With the slit width selected
to be 1.0 arcsec, the total spectral coverage was
3500 -- 6300 \AA\ and FWHM resolution 4.7 \AA\
(after $2 \times 2$ on-chip binning).
Table \ref{optical_setup} summarises the relevant 
instrument setup information.  
  
A total of 17.2 hours on the VLT was distributed between Periods
75 and 76 (see Table \ref{tel_time}).
All observations were carried out in service
mode.  Observing conditions were set according to
the target's optical magnitude.  Typically, the airmass
was less than 1.5, the seeing better than 1.0 arcsec and
FLI $<$ 50\%.

All calibration frames were obtained on the morning following
the science observations.  The flats were obtained using
an internal lamp and wavelength calibrations with a HgCd
lamp.  Extensive study of the stability of FORS2 indicates
that there is no significant wavelength drift over this
timescale.

\section{Data reduction}\label{data_sec}

The data obtained at the WHT and the VLT were reduced using tasks within the
IRAF package.  
A master bias was constructed from typically 5 -- 10
individual frames that were averaged together with
a high pixel and cosmic ray rejection algorithm within the IRAF 
task \textit{zerocombine}.   

Flat fields at both the WHT and VLT were obtained from internal
calibration lamps.  In both cases, typically 5 exposures were taken.
Each frame was visually inspected for potential saturation.  Acceptable
frames were averaged together with
a cosmic ray rejection algorithm within the IRAF task 
\textit{flatcombine}.  The flats were normalized to remove any 
intrinsic spectral shape and the strong vignetting at the ends of the 
ISIS frames.  There was no evidence for other strong detector- or 
instrument-induced effects, such as fringing.

All of the science frames were processed through the IRAF task 
\textit{ccdproc} in order to fit and trim the CCD overscan, subtract 
the master bias and divide through  by the master flat.

Cosmic rays were removed from the 2D frames with a custom
written software (based on IRAF routines) for longslit spectra.
First, the task \textit{background} is used to fit the columns
(spatial axis) of the 2D science (QSO) images.  This fit is
subtracted from the original 2D frame, effectively removing
the sky \footnote{This simple procedure only works well in the
absence of significant curvature.  Since we are dealing with
point sources, and instruments whose curvature is negligible over
at least $\pm$ 200 pixels either side of the QSO, this technique
works well.}.  A median filter is then applied along the spectral
axis, which removes the QSO signal.  The resulting
2D frame has now been cleaned of both sky and QSO, leaving behind 
a cosmic ray image.  A cosmic ray mask is then constructed by
setting all cosmic ray pixels to have a value of 10000 and all other pixels
to zero (in practice, this procedure
is achieved in multiple steps, to identify both the cosmic ray
`core' and wings).  The IRAF task \textit{fixpix} is used to apply 
the mask to the science images.

The QSO spectra were extracted from the 2D images using the task
\textit{apall} with optimal extraction.  The 1D spectra were
wavelength calibrated by applying a polynomial solution derived
from arc exposures obtained either during the night (WHT) or
as part of  the VLT daytime calibration plan.
The arcs were processed through the \textit{identify} task in IRAF,
with the final wavelength solution applied using \textit{dispcor}.
A polynomial fit was applied to derive the final wavelength solution;
a third -- fifth order Legendre polynomial was usually required, which
yielded typical RMS values $<$0.1 \AA.
As noted in the previous section, there can be significant drift
in the wavelength calibration of the ISIS instrument during the
night. We therefore observed each target with the following sequence:
arc, science exposure~1, science exposure~2, arc.  During the calibration
process, the arc taken immediately adjacent to the science exposure was
given a 2/3 weighting; the remaining arc was given a 1/3 weighting.

The GMOS reductions were executed using the Gemini IRAF
package.  This software operates within the IRAF infrastructure,
but has been specifically designed to process data from
Gemini instruments.  All frames are run through the task
\textit{gprepare}, adding appropriate header information.
The master bias frame was provided directly by Gemini
science operations. Five flat field frames were obtained from
observations of an internal calibration lamp.  All were
inspected for signs of saturation before constructing a master
flat using the task \textit{gsflat}.  The flats were averaged
together and a cosmic ray rejection algorithm used before
the master flat was normalised.  No significant fringing
was present, due to the relatively blue wavelength coverage
of our spectra.  After this initial processing, the data
were copied out of the Gemini multi-extension fits format,
so that the remainder of the data reduction could be
executed with standard IRAF tasks.

The rest of the GMOS data reduction proceeded as per
the WHT and VLT processing.  The 1D QSO spectra were
extracted using optimal weighting in \textit{apall}.
A trace of the QSO position on the CCD was used to extract
1D arc spectra, which were obtained as observations interleaved
with science exposures.  Wavelength calibration was done
using \textit{identify} and \textit{dispcor}.  As for the
WHT and VLT spectra, a third -- fifth order Legendre polynomial was 
usually required to derive the wavelength solution, 
yielding typical RMS values $<$0.1 \AA.

The wavelength calibrated 1D spectra were median combined using
\textit{scombine} with
equal weightings.  No flux calibration was attempted.
The spectra of all of the QSOs observed in this survey are
shown in Figure 1.

\subsection{Notes on individual objects}

\noindent \textbf{TXS0351+187} is a broad absorption line (BAL) quasar. 
The broad feature at 
$\lambda_{\rm obs}$ = 3750 \AA\ that resembles a DLA is Ly$\beta$ 
associated with the BAL outflow.\\

\smallskip

\noindent \textbf{TXS1013+235} is a BAL quasar.\\

\smallskip

\noindent \textbf{TXS2034+046} is reported in Dodonov et al. (1999) 
to have an emission redshift $z_{\rm em}=2.95$.  Our spectrum shows
Mg\,{\sc ii} $\lambda\lambda$2796, 2803 and C\,{\sc iii}]
$\lambda$1909, revealing that the true redshift is $z_{\rm em}=0.7165$.
This QSO is therefore not suitable for our DLA survey and we do
not include it in our main sample.  We present the spectrum of
this QSO in the Appendix and in Figure \ref{2034}\\

\smallskip

In addition to the quasars listed in Table \ref{optical_obs}, we
also attempted to 
obtain spectra of TXS1713$+$218 ($z_{\rm em} = 4.01$, $B$=22.0)
and TXS1835$-$345 ($z_{\rm em} = 2.78$, $B$=21.5), both of which
were observed at the VLT in service mode. Our spectrum of
TXS1713$+$218 has very little flux bluewards of Lyman-$\alpha$ emission,
so these data are not suitable for a DLA search. The spectrum of 
TXS1835$-$345 is clearly that of
a late-type star and inspection of the acquisition images confirms
that the wrong object was observed.  We therefore exclude these 
two targets from our tables and figures.

\clearpage
\begin{figure*}
\centerline{\rotatebox{0}{\resizebox{16cm}{!}
{\includegraphics{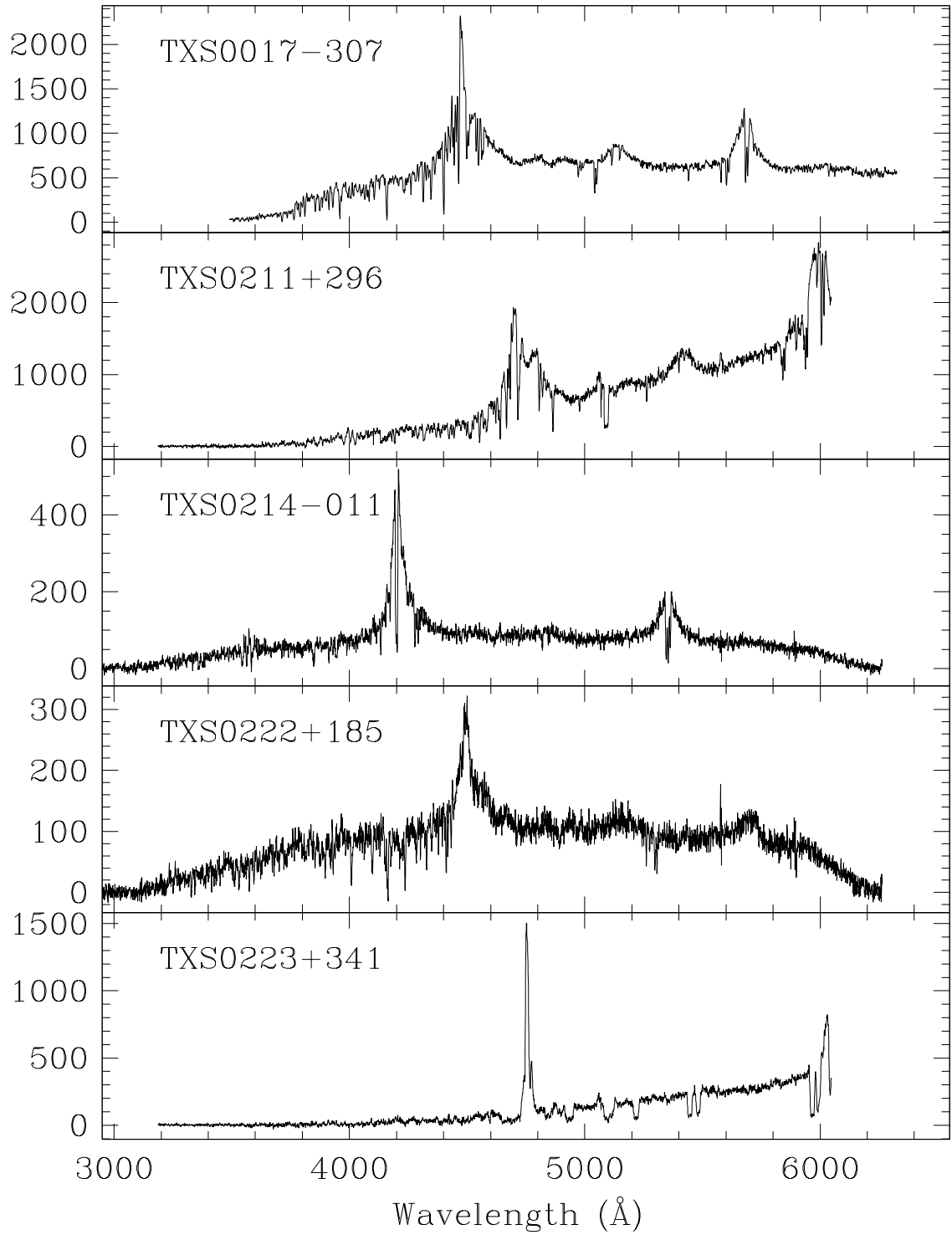}}}}
\caption{Spectra of the QSOs observed in our survey. \label{qsofig_1} }
\end{figure*}

\clearpage
\setcounter{figure}{0}
\begin{figure*}
\centerline{\rotatebox{0}{\resizebox{16cm}{!}
{\includegraphics{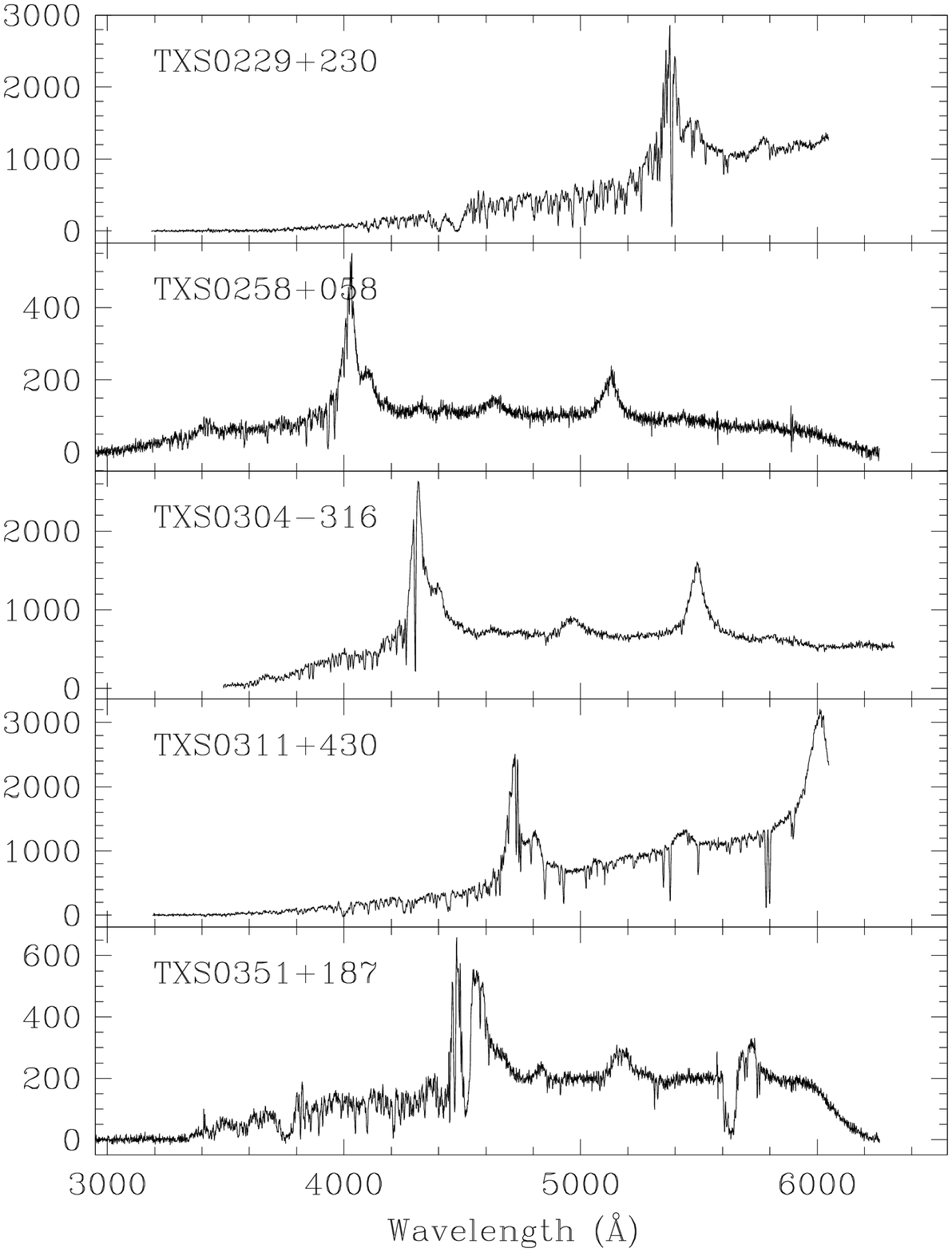}}}}
\caption{Continued.\label{qsofig_2} }
\end{figure*}

\clearpage
\setcounter{figure}{0}
\begin{figure*}
\centerline{\rotatebox{0}{\resizebox{16cm}{!}
{\includegraphics{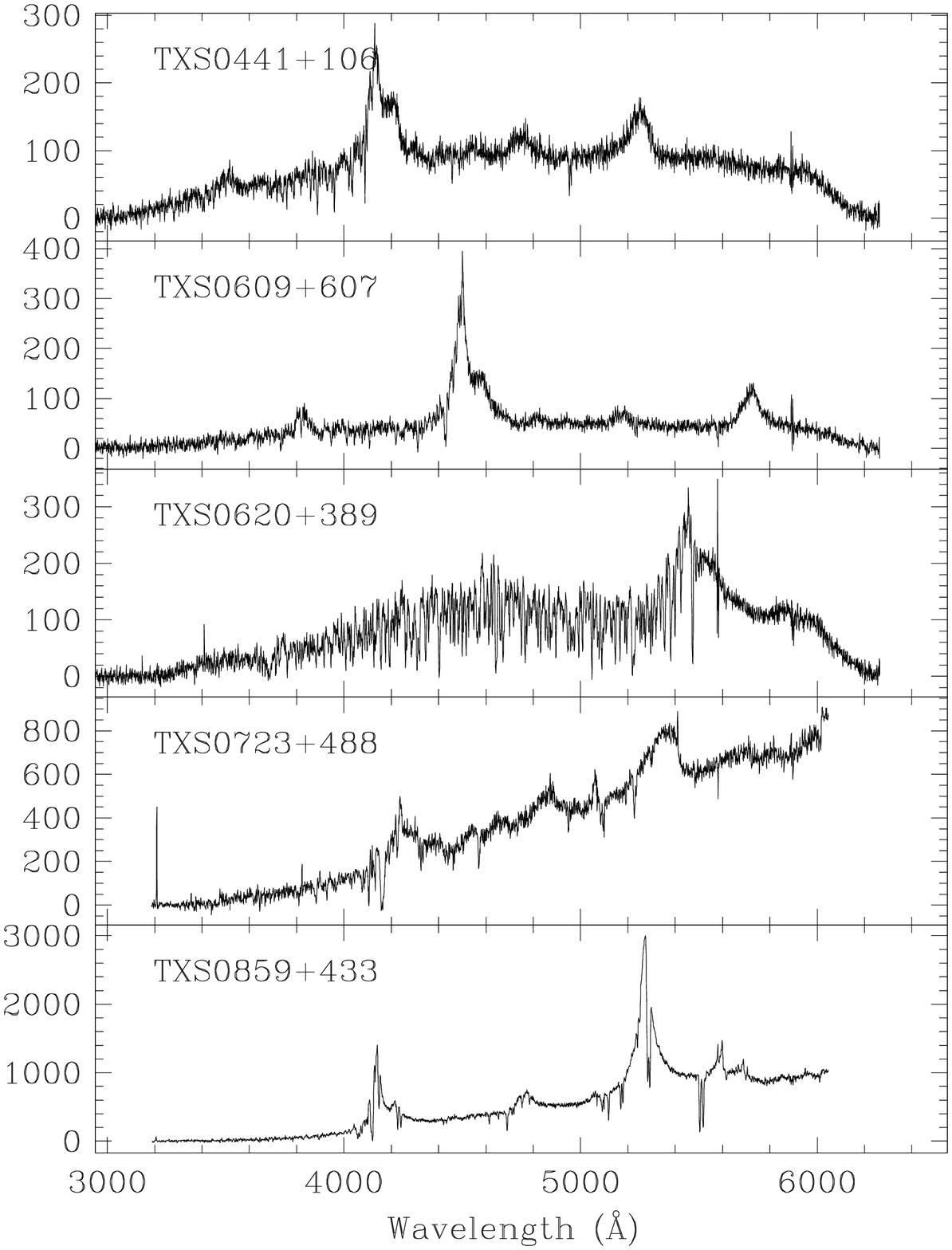}}}}
\caption{Continued.\label{qsofig_3} }
\end{figure*}

\clearpage
\setcounter{figure}{0}
\begin{figure*}
\centerline{\rotatebox{0}{\resizebox{16cm}{!}
{\includegraphics{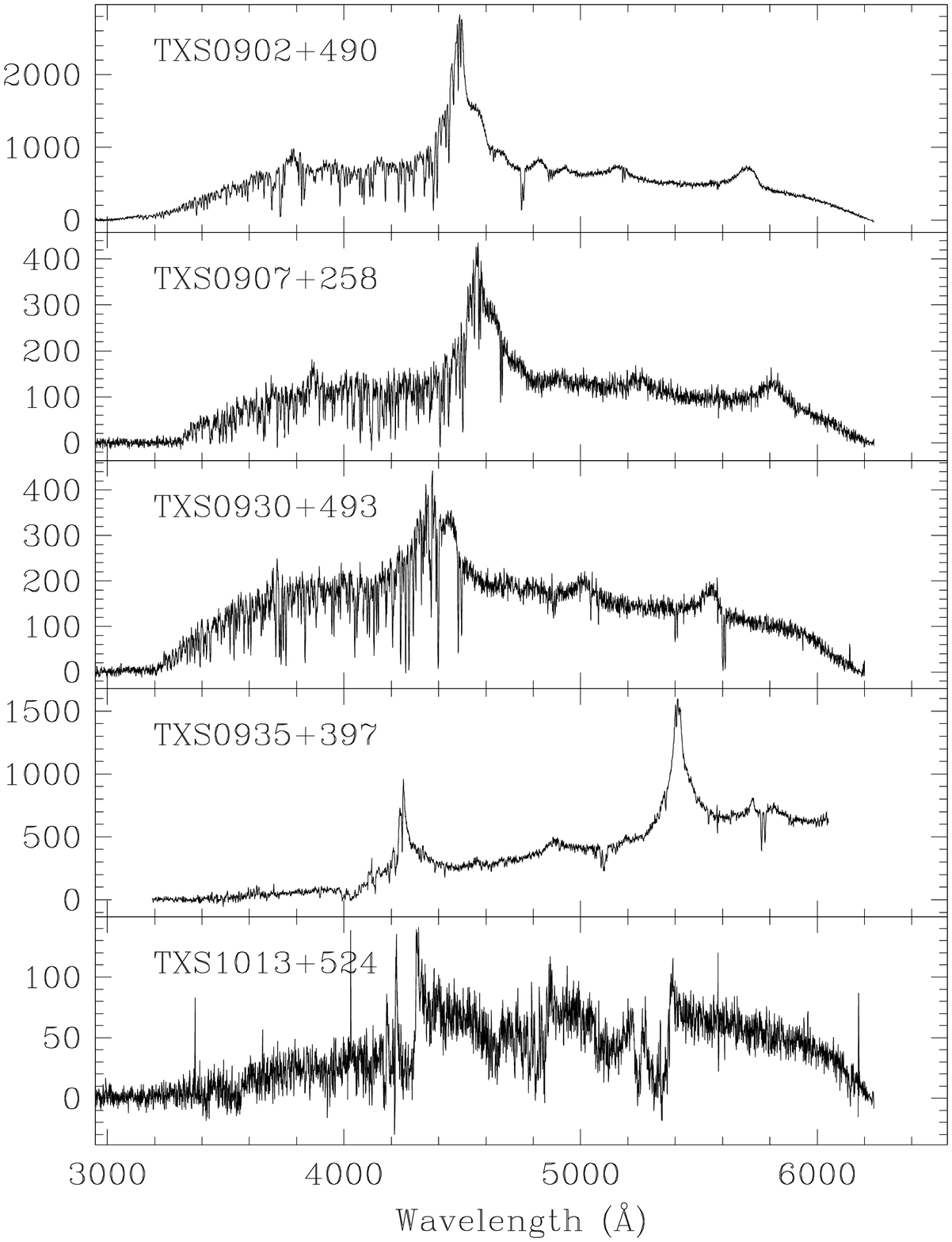}}}}
\caption{Continued.\label{qsofig_4} }
\end{figure*}

\clearpage
\setcounter{figure}{0}
\begin{figure*}
\centerline{\rotatebox{0}{\resizebox{16cm}{!}
{\includegraphics{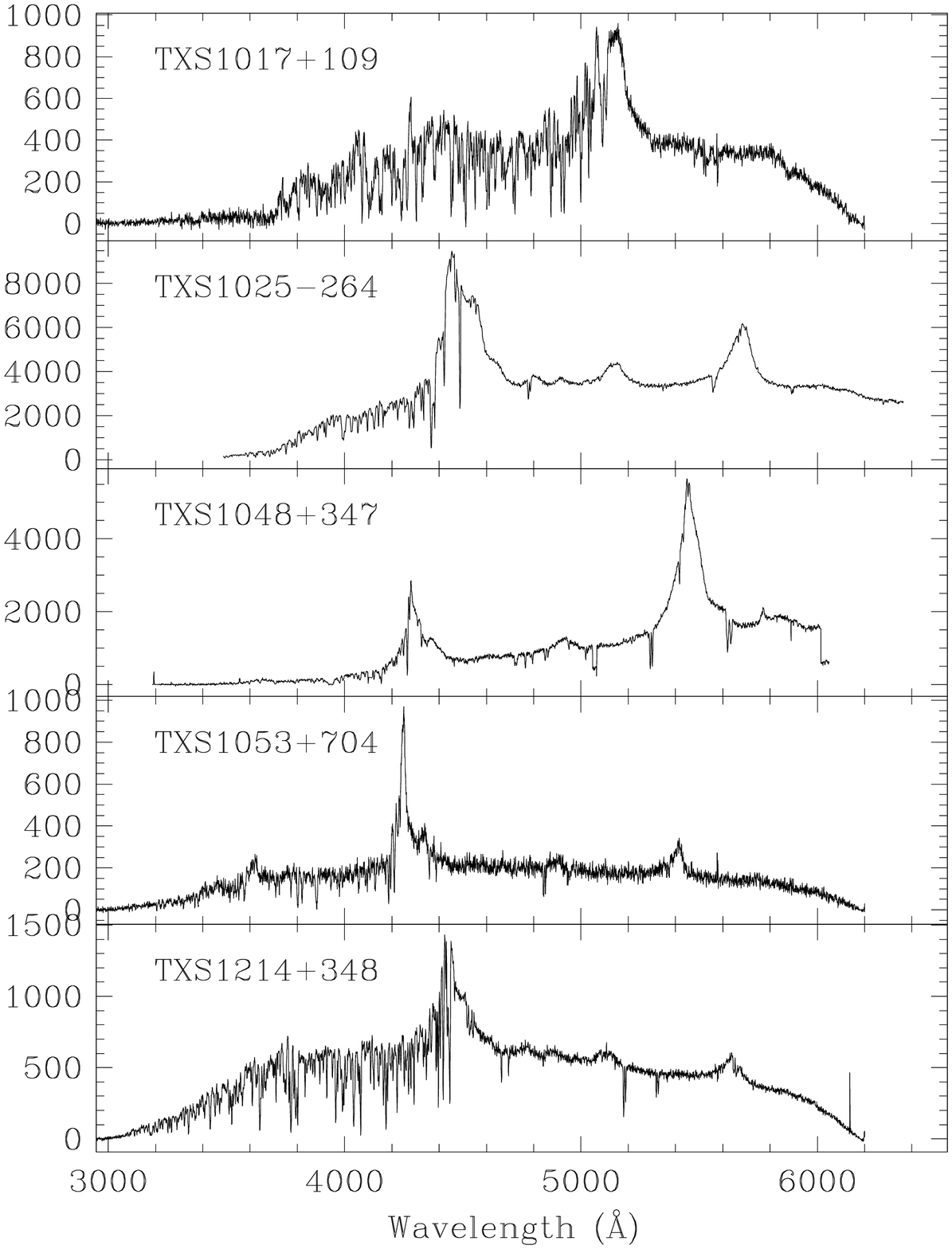}}}}
\caption{Continued.\label{qsofig_5} }
\end{figure*}

\clearpage
\setcounter{figure}{0}
\begin{figure*}
\centerline{\rotatebox{0}{\resizebox{16cm}{!}
{\includegraphics{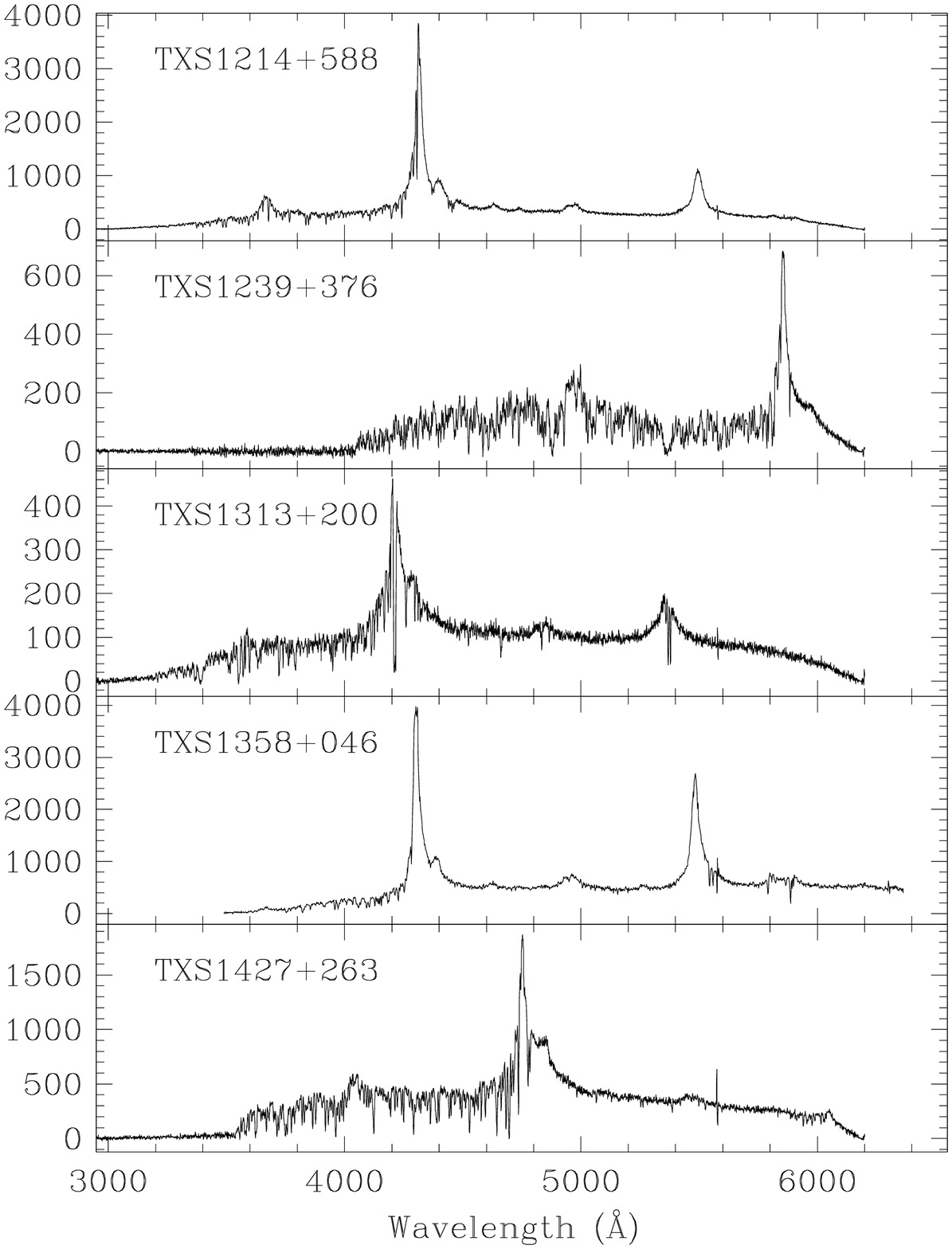}}}}
\caption{Continued.\label{qsofig_6} }
\end{figure*}

\clearpage
\setcounter{figure}{0}
\begin{figure*}
\centerline{\rotatebox{0}{\resizebox{16cm}{!}
{\includegraphics{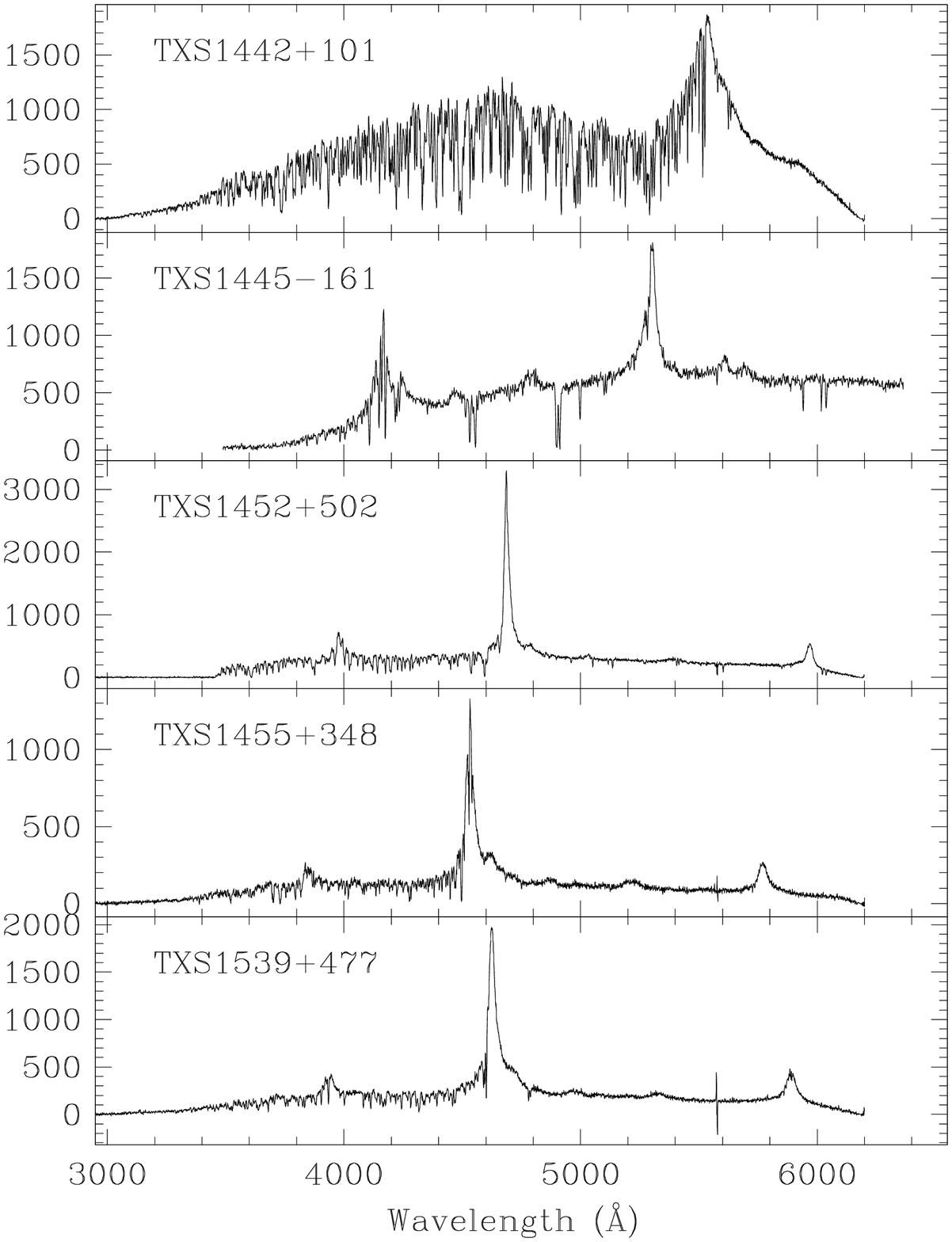}}}}
\caption{Continued.\label{qsofig_7} }
\end{figure*}

\clearpage
\setcounter{figure}{0}
\begin{figure*}
\centerline{\rotatebox{0}{\resizebox{16cm}{!}
{\includegraphics{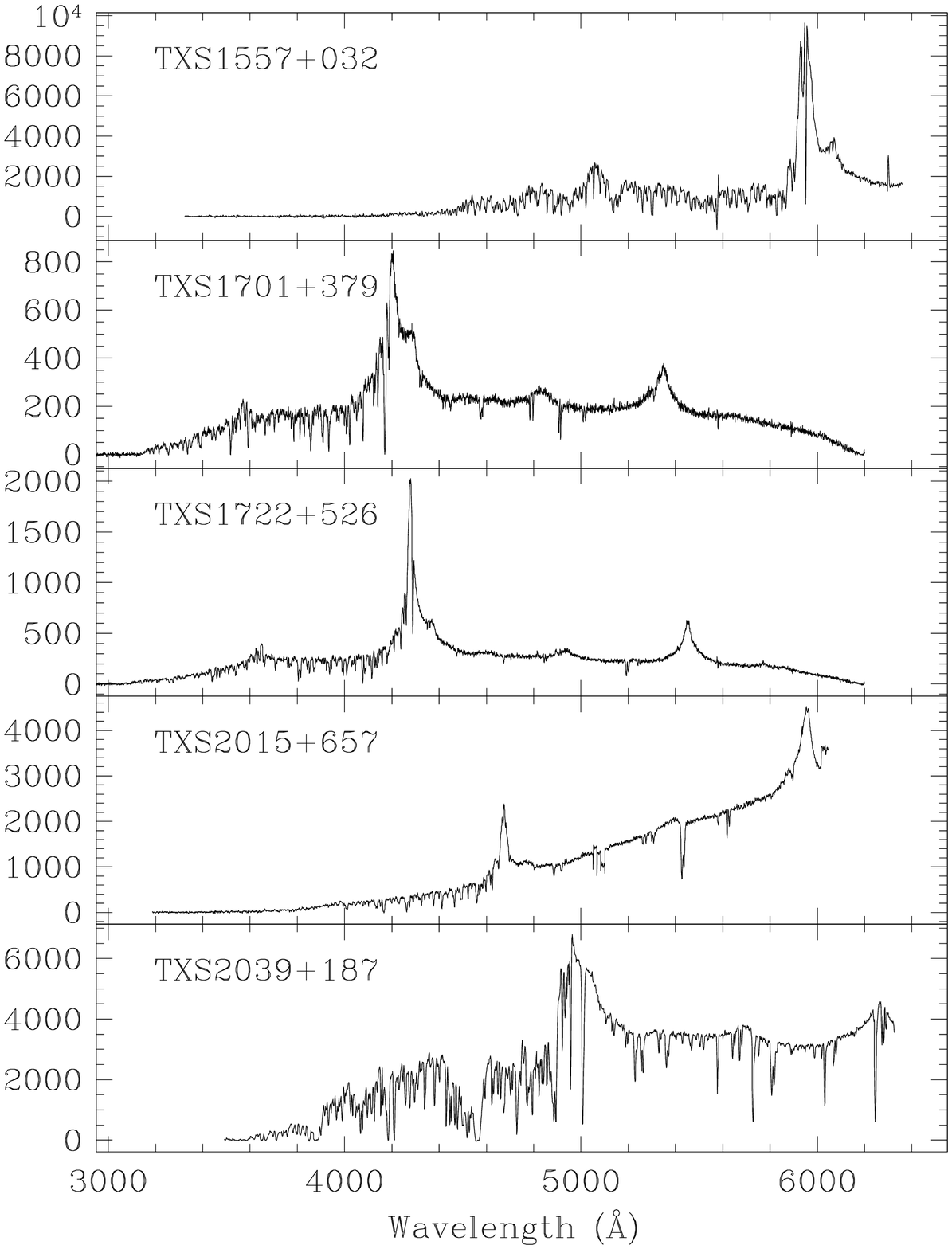}}}}
\caption{Continued.\label{qsofig_8} }
\end{figure*}

\clearpage
\setcounter{figure}{0}
\begin{figure*}
\centerline{\rotatebox{0}{\resizebox{16cm}{!}
{\includegraphics{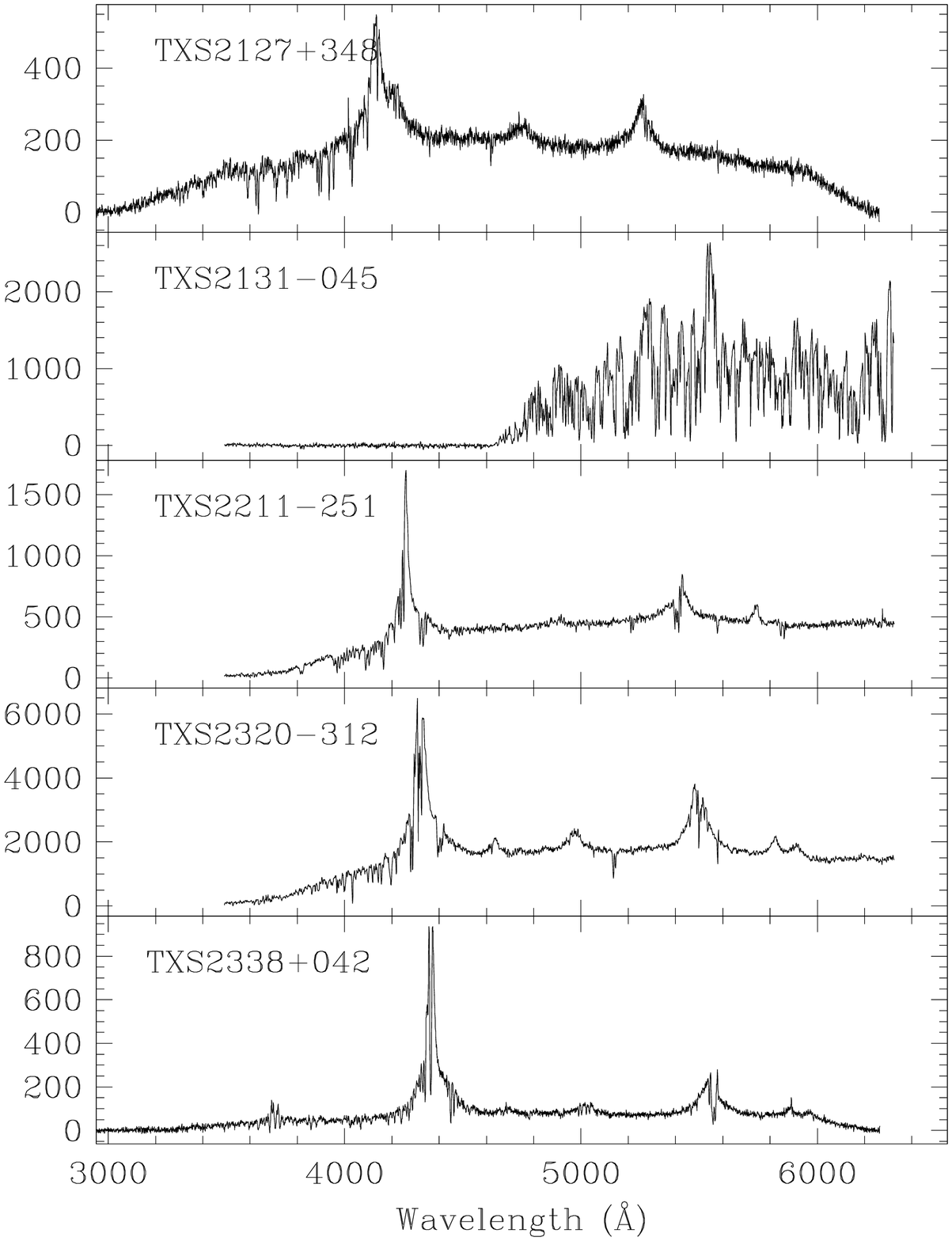}}}}
\caption{Continued.\label{qsofig_9} }
\end{figure*}
\clearpage

\section{Results}\label{dla_sec}

\setcounter{figure}{1}
\begin{figure}
\centerline{\rotatebox{0}{\resizebox{9cm}{!}
{\includegraphics{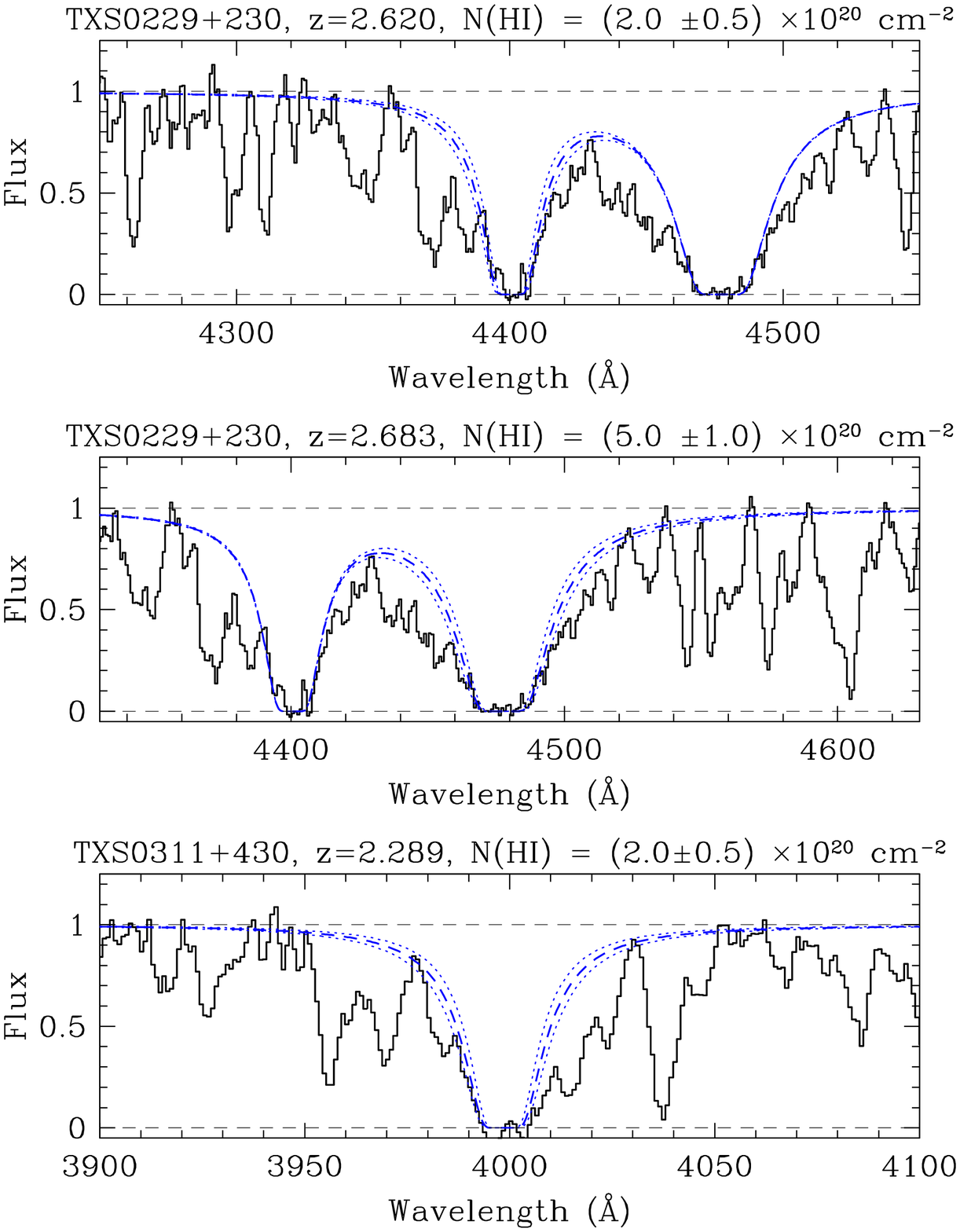}}}}
\caption{Profile fits to detected DLAs.  There are two DLAs towards
TXS0229+230, so the fit for each accounts for the damping wings
of the other.  \label{dla1} }
\end{figure}

\setcounter{figure}{1}
\begin{figure}
\centerline{\rotatebox{0}{\resizebox{9cm}{!}
{\includegraphics{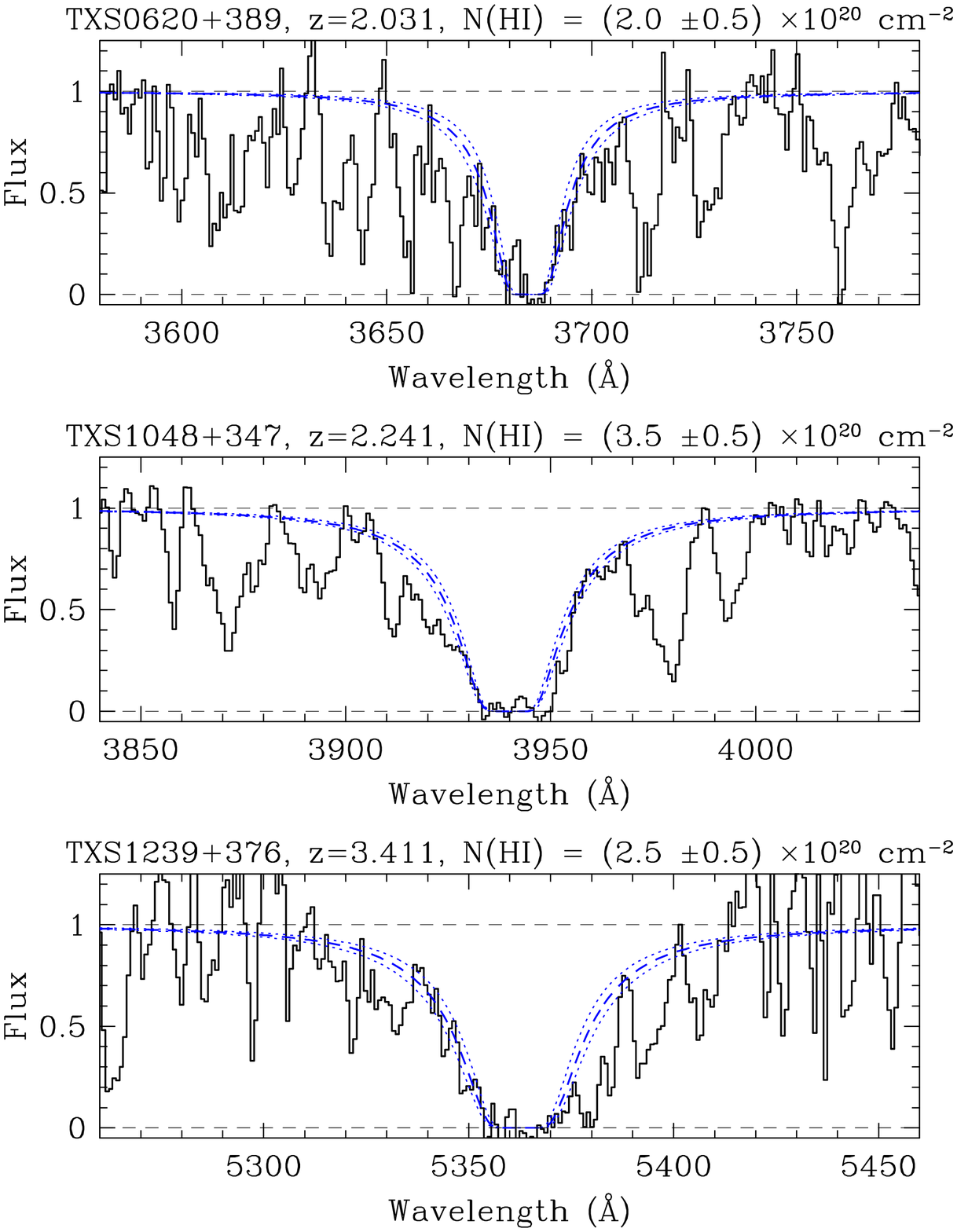}}}}
\caption{Continued. The residual flux in the Lyman-$\alpha$ 
trough of TXS0620+389 is almost certainly spurious,
since it appears in only one of the two exposures.  It may be caused,
at least in part, by a cosmic ray event very close the spectrum
on the CCD that was not fully removed. \label{dla2}}
\end{figure}

\setcounter{figure}{1}
\begin{figure}
\centerline{\rotatebox{0}{\resizebox{9cm}{!}
{\includegraphics{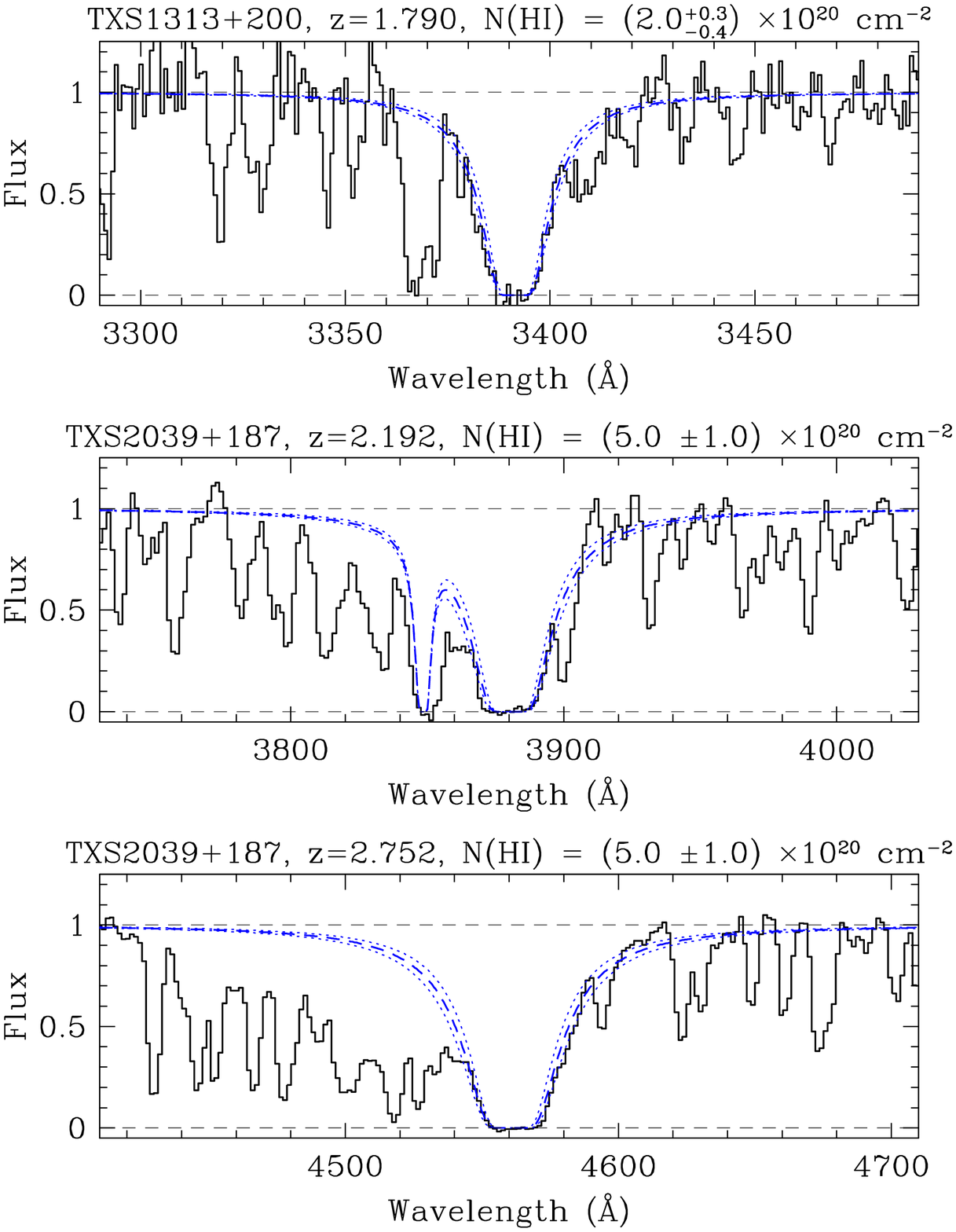}}}}
\caption{Continued.  The Ly$\beta$ transition
of the $z_{\rm abs}$= 2.752 DLA towards TXS2039+187 is accounted for
in the fit of the $z_{\rm abs} = 2.192$ DLA. \label{dla3} }
\end{figure}

\begin{figure}
\centerline{\rotatebox{270}{\resizebox{7cm}{!}
{\includegraphics{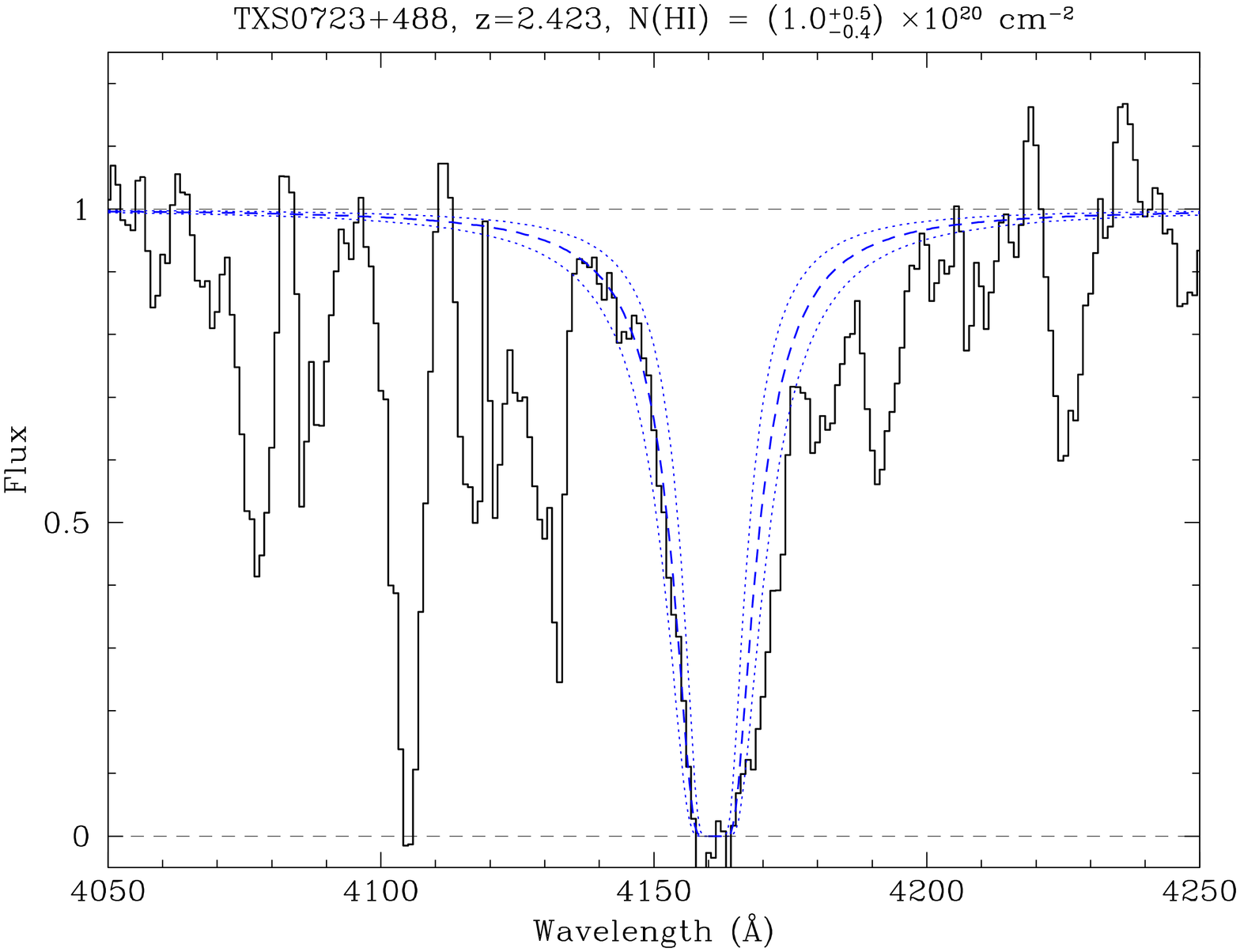}}}}
\caption{Profile fit to the detected sub-DLA.\label{sdla} }
\end{figure}

\subsection{DLA detection}

The spectra were visually inspected for DLA candidates.
We searched for absorbers that were separated
from the emission redshift of the QSO by at least
3000 \kms.  This velocity cut-off (sometimes extended
to 5000 \kms) has been a standard
custom in DLA searches, largely to exclude absorption
that may be associated with the QSO itself.
However, further motivation for a velocity cut-off
has been provided by Ellison et al. (2002) who
found an excess of absorbers within 3000 \kms\
of a radio-loud QSO sample.  Similar excesses
have now also been confirmed towards radio-quiet
QSOs (Russell, Ellison \& Benn 2006; Prochaska,
Hennawi \& Herbert-Fort 2008), indicating that these
`proximate' absorbers may be galaxies clustered
around the quasar\footnote{In any case, no proximate
DLAs are present in our survey.}.  
The maximum redshift at which we search for DLAs is therefore 

\begin{equation}
z_{\rm max} = \left[ (z_{\rm em} + 1) \sqrt{ \frac{(1-b)}{(1+b)} } \right] - 1,
\end{equation}

\noindent where

\begin{equation}
b = \frac{v}{c} = 0.01.
\end{equation}

The minimum redshift at which a DLA can be
identified is usually set by the blue wavelength
coverage of the spectrum (see Table \ref{optical_setup}).
However, in cases where a Lyman limit system (LLS) is 
present in the QSO spectrum within the redshift 
search range,

\begin{equation}
z_{\rm min} = \frac{(z_{\rm LLS} + 1) \times 912}{1216} - 1.
\end{equation}

The total redshift path for the survey is then

\begin{equation} 
\Delta z = \sum_{i = 1}^{n} (z_{i,\rm max} - z_{i,\rm min}) .
\end{equation}

For our survey of DLAs towards TXS sources, we
cover a total redshift path of $\Delta z = 38.79$.  The redshift
path towards each individual quasar is given in
Table \ref{dla_table}.

DLA candidates were identified by their large
rest frame equivalent widths (always in excess of 5 \AA), saturated
absorption troughs and extended damping wings.
Once identified, the candidate was confirmed by fitting
the continuum around the absorber, normalizing the
spectrum and comparing the absorption troughs
with theoretical damped profiles.  We identified a
total of nine DLAs with \nhi\ $\ge$ $2 \times 10^{20}$ \cm, plus one
absorber with \nhi\ = $1.0 \times 10^{20}$ \cm.  This latter
system is technically a sub-DLA and we therefore exclude it
from statements that refer specifically to DLA statistics.
The Lyman-$\alpha$ line of the 10 absorbers and theoretical damped profiles
are shown in Figures \ref{dla1} and \ref{sdla}.
One of the DLAs in our sample, towards TXS1239+376,
has also  been
identified in the survey of Jorgenson et al. (2006), although
this was not known at the time of our observations in April 2005.
Jorgenson et al. (2006) determine a column density of log \nhi\ = 
20.30$\pm$0.15; this value is slightly lower than our value of log \nhi\ = 
20.40$^{+0.08}_{-0.10}$, although consistent within the errors.
The QSO TXS0620+389 was also included in the Jorgenson et al.
survey, yet these authors do not report the detection of
a DLA in this sightline.  The reason for this omission is that
the rest-frame equivalent width measured for this absorber in the UCSD survey
was $<$5 \AA, and it therefore did not pass the
threshold to be considered as a candidate DLA
(R. Jorgenson, private communication). 

From the sample of nine DLAs, we derive a number density
$n(z) = \# DLA / \Delta z = 0.23^{+0.11}_{-0.07}$,  
where the error bars are derived from the 1 $\sigma$ confidence limits
applicable to small numbers, as tabulated by Gehrels (1986). 
The mean absorption redshift of our nine DLAs is $\langle z \rangle = 2.44$, 
a value
that decreases slightly to $\langle z \rangle = 2.40$ if the single 
$z_{\rm abs} < 2$
and $z_{\rm abs} > 3$ DLAs are excluded.  Our derived number density
is therefore in good agreement with the functional fit to the SDSS-DR3 
DLA sample (Prochaska, Herbert-Fort \& Wolfe 2005):  Eqn.~5 in 
Prochaska, Hennawi \& Herbert-Fort (2008) yields $n(z) = 0.19$ for
$\langle z \rangle = 2.44$.  For comparison, the CORALS survey
covered $\Delta z = 57.16$ and detected 19 DLAs at a mean absorption
redshift of $\langle z \rangle = 2.50$ (Ellison et al. 2001).
The UCSD survey covered $\Delta z = 41.15$ and detected 7 DLAs at a 
mean absorption redshift of $\langle z \rangle = 2.53$ 
(Jorgenson et al. 2006).  The combined statistics of these two surveys
yields $n(z) = 0.26^{+0.06}_{-0.05}$ (Jorgenson et al. 2006),
indicating that the number density can be fairly robustly determined,
even from relatively small samples.  Unlike the
CORALS and UCSD survey, our sample is not optically complete.  The
consistency of our number density with the both the
SDSS-DR3 sample and the radio-selected CORALS and UCSD samples is
further evidence that dust depletion is unlikely to significantly affect
the derivation of $n(z)$ from magnitude-limited, optically-selected
QSO samples (Ellison et al. 2001; Jorgenson et al. 2006).

\begin{center}
\begin{table*}
\caption{Absorber search statistics}
\begin{tabular}{lccccc}
\hline
QSO & $z_{\rm max}$ & $z_{\rm min}$ & $\Delta z$ &$z_{\rm abs}$ & \nhi\\\
    &               &               &           &          &  (10$^{20}$ \cm)\\
\hline
TXS0017$-$307  &  2.63  & 1.90  &  0.73  & ... & ... \\ 
TXS0211$+$296  &  2.83  & 2.04  &  0.79  & ... & ... \\ 
TXS0214$-$011  &  2.42  & 1.63  &  0.79  & ... & ... \\ 
TXS0222$+$185  &  2.65  & 1.63  &  1.02  & ... & ... \\ 
TXS0223$+$341  &  2.87  & 2.45  &  0.42  & ... & ... \\ 
TXS0229$+$230  &  3.38  & 2.06  &  1.32  & 2.620 $\pm$ 0.001   &   2.0 $\pm$ 0.5\\
~~~~~~~~~~~~~  &  ~~~~  & ~~~~  &  ~~~~  & 2.683 $\pm$ 0.002   &   5.0 $\pm$ 1.0\\
TXS0258$+$058  &  2.28  & 1.59  &  0.69  & ... & ... \\ 
TXS0304$-$316  &  2.50  & 1.90  &  0.60  & ... & ... \\ 
TXS0311$+$430  &  2.83  & 2.04  &  0.79  & 2.289 $\pm$ 0.002   &   2.0 $\pm$ 0.5\\
TXS0351$+$187  &  2.67  & 1.75  &  0.92  & ... & ... \\ 
TXS0441$+$106  &  2.37  & 1.63  &  0.74  & ... & ... \\ 
TXS0609$+$607  &  2.66  & 1.80  &  0.86  & ... & ... \\ 
TXS0620$+$389  &  3.42  & 1.71  &  1.71  &  2.031 $\pm$ 0.002   &   2.0 $\pm$ 0.5\\
TXS0723$+$488  &  2.43  & 1.96  &  0.47  &  2.423 $\pm$ 0.002   &   1.0 $^{+0.5}_{-0.4}$\\
TXS0859$+$433  &  2.38  & 2.04  &  0.34  & ... & ... \\ 
TXS0902$+$490  &  2.65  & 1.63  &  1.02  & ... & ... \\ 
TXS0907$+$258  &  2.70  & 1.75  &  0.95  & ... & ... \\ 
TXS0930$+$493  &  2.54  & 1.67  &  0.87  & ... & ... \\ 
TXS0935$+$397  &  2.46  & 2.04  &  0.42  & ... & ... \\ 
TXS1013$+$524  &  2.42  & 1.96  &  0.46  & ... & ... \\ 
TXS1017$+$109  &  3.11  & 2.06  &  1.05  & ... & ... \\ 
TXS1025$-$264  &  2.62  & 1.90  &  0.72  & ... & ... \\ 
TXS1048$+$347  &  2.48  & 1.96  &  0.52  &  2.241 $\pm$ 0.002   &   3.5 $\pm$ 0.5\\
TXS1053$+$704  &  2.46  & 1.63  &  0.83  & ... & ... \\ 
TXS1214$+$348  &  2.60  & 1.59  &  1.01  & ... & ... \\ 
TXS1214$+$588  &  2.50  & 1.59  &  0.91  & ... & ... \\ 
TXS1239$+$376  &  3.76  & 2.37  &  1.39  &  3.411 $\pm$ 0.003   &   2.5 $\pm$ 0.5\\
TXS1313$+$200  &  2.44  & 1.63  &  0.81  &  1.790 $\pm$ 0.002   &   2.0 $^{+0.3}_{-0.4}$\\
TXS1358$+$046  &  2.51  & 1.90  &  0.61  & ... & ... \\ 
TXS1427$+$263  &  2.87  & 1.96  &  0.91  & ... & ... \\ 
TXS1442$+$101  &  3.48  & 1.63  &  1.85  & ... & ... \\ 
TXS1445$-$161  &  2.38  & 2.00  &  0.38  & ... & ... \\ 
TXS1452$+$502  &  2.80  & 1.88  &  0.92  & ... & ... \\ 
TXS1455$+$348  &  2.69  & 1.63  &  1.06  & ... & ... \\ 
TXS1539$+$477  &  2.76  & 1.59  &  1.17  & ... & ... \\ 
TXS1557$+$032  &  3.84  & 2.50  &  1.34  & ... & ... \\ 
TXS1701$+$379  &  2.42  & 1.63  &  0.79  & ... & ... \\ 
TXS1722$+$526  &  2.48  & 1.59  &  0.89  & ... & ... \\ 
TXS2015$+$657  &  2.80  & 2.04  &  0.76  & ... & ... \\ 
TXS2039$+$187  &  3.01  & 2.00  &  1.01  &  2.192 $\pm$ 0.002   &   5.0 $\pm$ 1.0\\
~~~~~~~~~~~~   &  ~~~~  & ~~~~  &  ~~~~  &  2.752 $\pm$ 0.001   &   5.0 $\pm$ 1.0\\
TXS2127$+$348  &  2.37  & 1.59  &  0.78  & ... & ... \\ 
TXS2131$-$045  &  4.18  & 2.91  &  1.27  & ... & ... \\ 
TXS2211$-$251  &  2.47  & 1.92  &  0.55  & ... & ... \\ 
TXS2320$-$312  &  2.50  & 1.90  &  0.60  & ... & ... \\ 
TXS2338$+$042  &  2.55  & 1.80  &  0.75  & ... & ... \\ 
\hline 
\end{tabular}\label{dla_table}
\end{table*}
\end{center}

\subsection{Metals}

For all of the DLAs detected in our survey, we checked the spectra
for the presence of strong metal lines.  Although most metal lines
that can be detected at the S/N and resolution of our spectra are
likely to be saturated, they are useful for constraining the absorption
redshift of the DLA.  In some cases, they may also be used to obtain
a crude metallicity estimate.
Our spectroscopic observations were designed to maximise
coverage of the Lyman-$\alpha$ forest.  Since many of the strong metal lines that are
usually associated with DLAs are not within our spectral coverage,
we restricted our search to the following species: 
\fe2\ $\lambda$ 1608, \si2\ $\lambda$1304, \si2\ $\lambda 1526$, 
\si2\ $\lambda 1808$, \al2\ $\lambda 1670$.
In Table \ref{metals_ew} we give measured rest frame equivalent widths and
3 $\sigma$ limits in the case of non-detections.  The detection limits
were calculated on the assumption of a single unresolved component whose
FWHM matches the instrumental resolution (FWHM$_i$).  
In the rest frame of the absorber this 3 $\sigma$ equivalent width (EW) 
limit corresponds to 
\begin{equation}
EW(3\sigma) = \frac{3 \times FWHM_i}{S/N \times (1+z_{\rm abs})}
\end{equation}

Equivalent widths were determined by fitting one or multiple gaussians,
as well as direct integration across the entire absorption line.
This process is repeated for various estimates of the continuum level.
The quoted errors therefore reflect both the range of EWs that we 
measure from the two methods, and an estimate of the continuum error.
We compared the profiles of each metal line detection in velocity
space in order to assess the potential contamination of metal lines
by other species.  If the contaminating component is at least partially
resolved from the metal line of interest, we attempt to de-blend the components.
However, if the contamination is severe,
we do not report an EW and the entry `blended' appears in Table 
\ref{metals_ew}.  If the metal line is in the Lyman-$\alpha$ forest,
the entry `\lya' is given.

Although high resolution
(typically, echelle) spectra are normally required to
accurately measure DLA abundances, lower resolution spectra
have been used to determine metallicities from weak
absorption lines such as those of Zn\,{\sc ii} and Cr\,{\sc ii}
(e.g. Pettini et al. 1997).  However, high S/N ratios are required
to detect such weak lines, and our spectra typically have
detection limits that are too shallow for accurate metallicities
to be derived.  Nonetheless, we attempt a crude estimate of
the metallicities of our survey DLAs.  If any of the \si2\ or
\fe2\ lines have EWs $\le$ 0.20 \AA\ we use the assumption of
a linear curve of growth to determine the \si2\ or \fe2\
column density.  In practice, this is likely to be a lower
limit to the elemental column density 
if the line is partially saturated.  Alternatively, we use
the EW of the \si2\ $\lambda$1526 \AA\ transition as a rough metallicity
indicator, following the correlation of Prochaska et al. (2008):

\begin{equation}\label{metal_eqn}
[M/H] = (-0.92 \pm 0.05) + (1.41 \pm 0.10) \log {\rm EW (Si~II~ \lambda 1526)}
\end{equation}

Although the \textit{mean} value of metallicity for a given
EW (Si\,{\sc ii} ~$\lambda 1526$) is well determined from this
calibration, often to better than 0.1 dex according to Prochaska
et al. (2008), there is nevertheless a large spread in the metallicities
of DLAs about the relation in Eqn \ref{metal_eqn}. The metallicities
of individual DLAs derived from this technique should therefore
be treated with caution, and as rough estimates only.

The metallicity estimates and the technique from which they
are derived are given in Table \ref{metals}.  In a number of
cases, the metallicity can be derived from both \si2\ and
\fe2\ lines.  In all such cases [Si/H] $>$ 
[Fe/H]\footnote{[X/H]=log(N(X)/\nhi)$-$log(N(X)/\nhi)$_{\odot}$. 
Solar values from Lodders (2003).}, as is
usually the case in DLAs due to preferential dust depletion
of iron, and/or an enhancement
of alpha-capture elements such as silicon.  The metallicities that
we obtain are mostly in the range of 1/30 to 1/100 of the
solar value, typical of other DLAs at these redshifts 
(e.g. Prochaska et al.  2003).  The three exceptions are the DLAs towards
TXS0311+430, TXS1313+200 and the $z_{\rm abs} = 2.752$ absorber
towards TXS2039+187.  Taking into account
the two estimates of [Si/H] in the DLA towards TXS0311+430,
and acknowledging that the curve of growth value may actually
be a lower limit, this DLA appears to have a metallicity 
of at least 1/16, and possibly as high 
as 1/4, of the solar value\footnote{Our \si2\ curve of growth abundance,
[Si/H] = $-0.6$ is slightly lower than the value derived by York et al. 
(2007), who found [Si/H] = $-0.48$.  This is due to a re-assessment
of the continuum level.}.  This DLA has also been found to be
the only known case to date of a high redshift DLA with low
spin temperature, T$_{\rm s} = (138 \pm 36)$~K (York et al. 2007), 
supporting the anti-correlation between metallicity and spin temperature 
tentatively detected in earlier DLA samples
(Kanekar \& Chengalur 2001; Kanekar \& Briggs 2004). 
If this is indeed the case, 
then the DLA towards TXS1313+200 and the $z_{\rm abs} = 2.752$ absorber
towards TXS2039+187 whose metallicities appear to be 
1/3 -- 1/10 and 1/3 of the solar value respectively, 
may also have low spin temperatures.  It will therefore be interesting
to confirm the unusually high metallicities of these DLAs with     
higher resolution spectroscopy. If they indeed arise in massive galaxies, 
as may be expected from the mass-metallicity relation observed in
high redshift galaxies (Savaglio et al. 2005; Erb et al. 2006; Liu et al. 
2008), Hubble Space Telescope images may be able to confirm their morphologies 
and luminosities.

\begin{center}
\begin{table*}
\caption{Metal line equivalent widths and 3$\sigma$ detection limits}
\begin{tabular}{lcccccc}
\hline
QSO & $z_{\rm abs}$ &  \fe2\ $\lambda$1608 &  \si2\ $\lambda$1304 &  \si2\ $\lambda$1526 &  \si2\ $\lambda$1808  &  \al2\ $\lambda$1670 \\ 
 & & (\AA) & (\AA) & (\AA) & (\AA) & (\AA) \\ \hline
TXS 0229+230 & 2.620 &  blended & \lya\ & 0.33$\pm$0.02 &  ... & ... \\
TXS 0229+230 & 2.683 & 0.13$\pm$0.02 &\lya\ & 0.36$\pm$0.05 & ... & ...\\
TXS 0311+430 & 2.289 & 0.21$\pm$0.02 & \lya\ & 0.69$\pm$0.05 & 0.10$\pm$0.01 & 0.73$\pm$0.03  \\
TXS 0620+389 & 2.031 & \lya\ & \lya\ & \lya\ & blended & \lya\ \\
TXS 0723+488 & 2.423 & $<$0.12 & 0.20$\pm$0.02 & blended & ... & 0.15$\pm$0.02 \\
TXS 1048+347 & 2.241 & 0.15$\pm$0.04& \lya\ & 0.26$\pm$0.06 & $<$0.09 & 0.25$\pm$0.01 \\
TXS 1239+376 & 3.411 & ... & \lya\ & ... & ... & ...\\
TXS 1313+200 & 1.790 & 0.23$\pm$0.05 & \lya\ & 0.89$\pm$0.03 & $<$0.18 & 0.97$\pm$0.05 \\
TXS 2039+187 & 2.192 & 0.17$\pm$0.03 & \lya\ & \lya\ & 0.04$\pm$0.01 & 0.38$\pm$0.02 \\
TXS 2039+187 & 2.752 & 1.12$\pm$0.03 & \lya\ & 1.90$\pm$0.02 & ... & 1.80$\pm$0.06 \\ \hline
\end{tabular}\label{metals_ew}
\end{table*}
\end{center}

\begin{center}
\begin{table}
\caption{DLA metallicity estimates}
\begin{tabular}{lcrl}
\hline
QSO & $z_{\rm abs}$ & [M/H] & Method \\  \hline
TXS 0229+230 & 2.620 & $-1.6$ &  \si2\ $\lambda$1526-[M/H] relation\\
TXS 0229+230 & 2.683 & $-2.2$ & \fe2\ $\lambda$1608 linear COG\\
~~~~~~~~~~~~ & ~~~~~ & $-1.6$ &  \si2\ $\lambda$1526-[M/H] relation\\
TXS 0311+430 & 2.289 & $-0.6$ & \si2\ $\lambda$1808 linear COG\\
~~~~~~~~~~~~ & ~~~~~ & $-1.2$ &  \si2\ $\lambda$1526-[M/H] relation\\
TXS 0620+389 & 2.031 & ... & N/A\\
TXS 0723+488 & 2.423 & $-1.4$ & \si2\ $\lambda$1304 linear COG\\ %
~~~~~~~~~~~~ & ~~~~~ & $<-1.5$ & \fe2\ $\lambda$1608 linear COG\\
TXS 1048+347 & 2.241 & $-1.7$ &  \si2\ $\lambda$1526-[M/H] relation\\
~~~~~~~~~~~~ & ~~~~~ & $<-0.9$& \si2\ $\lambda$1808 linear COG \\
~~~~~~~~~~~~ & ~~~~~ & $-2.0$ & \fe2\ $\lambda$1608 linear COG\\
TXS 1239+376 & 3.411 & ... & N/A\\
TXS 1313+200 & 1.790 & $<-0.4$ & \si2\ $\lambda$1808 linear COG\\
~~~~~~~~~~~~ & ~~~~~ & $-1.0$ &  \si2\ $\lambda$1526-[M/H] relation\\
TXS 2039+187 & 2.192 & $-1.4$ & \si2\ $\lambda$1808 linear COG\\
~~~~~~~~~~~~ & ~~~~~ & $-2.1$ & \fe2\ $\lambda$1608 linear COG\\
TXS 2039+187 & 2.752 & $-0.5$ & \si2\ $\lambda$1526-[M/H] relation  \\ \hline
\end{tabular}\label{metals}
\end{table}
\end{center}

\section{Summary}

We have presented the results of a survey for damped Lyman-$\alpha$
systems towards a sample of radio-selected QSOs.  The sample was
designed to identify DLAs that would be suitable for follow-up
absorption spectroscopy in the redshifted \hi\ 21\,cm line.  
Our sample consists of 45 QSOs, covering
a redshift path of $\Delta z$ = 38.79, towards which we detect
nine DLAs and one sub-DLA.  The DLA number density is therefore 
$n(z) = 0.23^{+0.11}_{-0.07}$ at a mean absorption redshift of $\langle z 
\rangle = 2.44$, in good agreement with previous surveys of both
optically- and radio-selected QSOs.
We present fits to the damped profiles and derive \hi\ column densities and measurements
of metal line equivalent widths for all detected DLAs.  We have already published an 
\hi\ 21\,cm detection towards one of these DLAs (TXS0311+430, $z_{\rm abs}
=2.289$, T$_{\rm s} = (138 \pm 36)$~K; York et al. 2007). In a future paper, Kanekar et
al. (in preparation) will report the rest of the results of our follow-up
\hi\ 21\,cm observations.

\section*{Acknowledgments} 
SLE was funded by an NSERC Discovery grant.  BAY was partially funded
by an NSERC PGS-M award. NK acknowledges support from the Max-Planck 
foundation.  The authors would like to thank the referee, Johan Fynbo,
for his useful and prompt review of this work.
This research has made use of the NASA/IPAC Extragalactic 
Database (NED) which is operated by the Jet Propulsion Laboratory, 
California Institute of Technology, under contract with the National 
Aeronautics and Space Administration.
Based on observations obtained at the Gemini Observatory, 
which is operated by the
Association of Universities for Research in Astronomy, Inc., 
under a cooperative agreement
with the NSF on behalf of the Gemini partnership: the National 
Science Foundation (United
States), the Science and Technology Facilities Council (United Kingdom), the
National Research Council (Canada), CONICYT (Chile), the Australian 
Research Council
(Australia), Ministrio da Cincia e Tecnologia (Brazil) 
and SECYT (Argentina).

\newpage

\begin{appendix}
\section{Updated redshift for TXS2034+046}

\begin{figure*}
\centerline{\rotatebox{270}{\resizebox{10cm}{16cm}
{\includegraphics{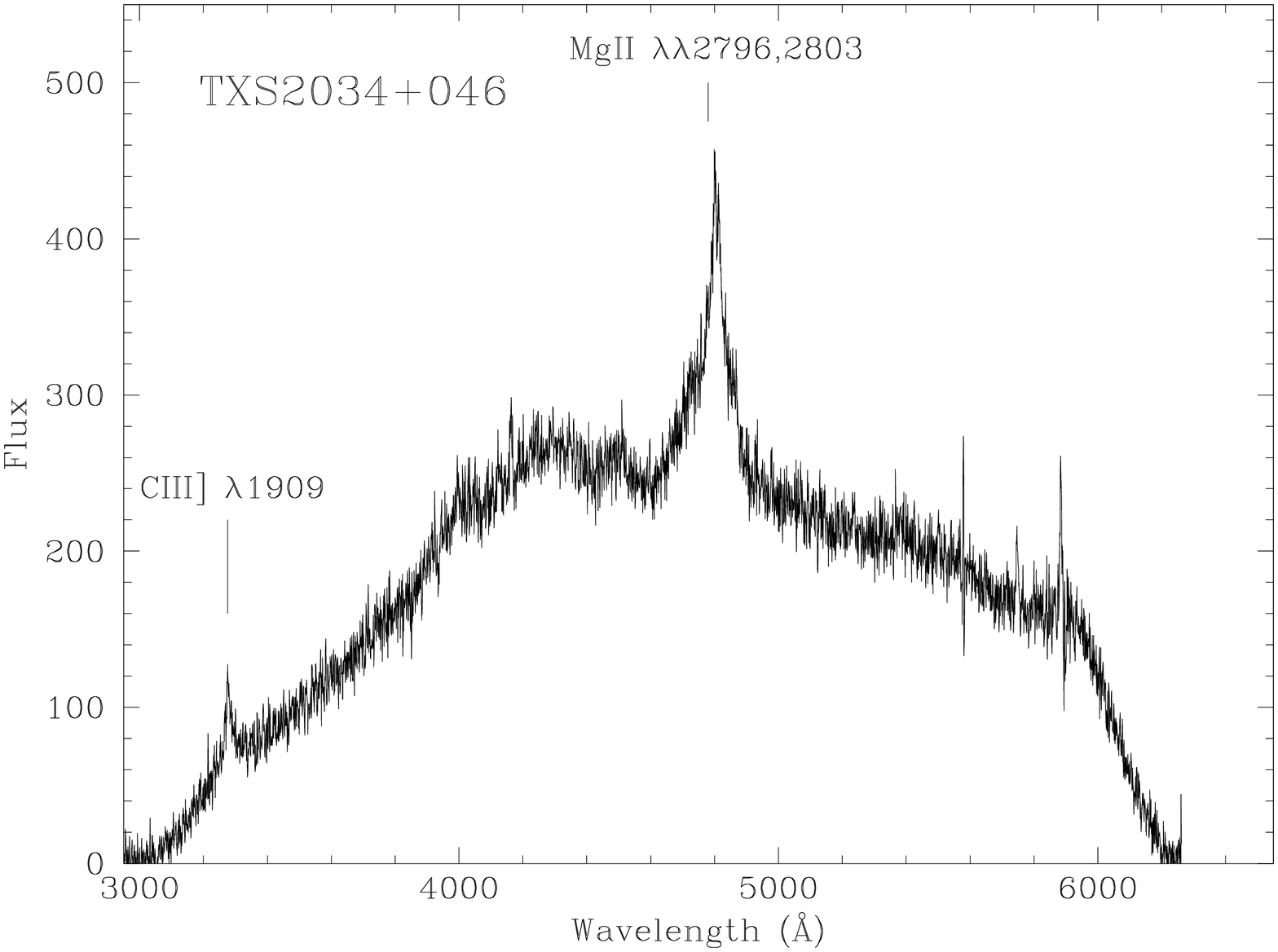}}}}
\caption{The spectrum of TXS2034+046, revealing its true redshift to
be $z_{\rm em}$=0.7165.\label{2034} }
\end{figure*}

We have obtained a spectrum of TXS2034+046 which 
 is reported in Dodonov et al. (1999) 
to have an emission redshift $z_{\rm em}=2.95$.  Figure \ref{2034}
shows the spectrum obtained at the WHT where emission lines corresponding
to Mg\,{\sc ii} $\lambda\lambda$2796, 2803
and C\,{\sc iii}]
$\lambda$1909 reveal that the true redshift is $z_{\rm em}$=0.7165.

\end{appendix}


\begin{thebibliography}{}

\bibitem[Braun (1997)]{bra97}
        Braun, R.,  1997, ApJ, 484, 637

\bibitem[Braun \& Walterbos (1992)]{bw92}
        Braun, R., \& Walterbos, R. A. M., 1992, ApJ 
	386, 120

\bibitem[Briggs, Brinks \& Wolfe (1997)]{bbw}
        Briggs, F.H., Brinks, E., \& Wolfe, A.M. 1997, AJ, 113, 467        

\bibitem[Carilli et al (1996)]{car96}
         Carilli, C. L., Lane, W., de Bruyn, A. G., Braun, R., 
	 Miley, G. K., 1996, AJ, 112, 1317

\bibitem[Chengalur \& Kanekar 2000]{ck00}
        Chengalur, J. N., \& Kanekar, N., 2000, MNRAS, 318, 303

\bibitem[Curran et al. (2005)]{cur05}
        Curran, S. J., Murphy, M. T., Pihlstrom, Y. M., Webb, J. K., 
	Purcell, P. R.,
	2005, MNRAS, 356, 1509

\bibitem[Curran \& Webb (2006)]{cw06}
        Curran, S. J., Webb, J. K.,  2006, MNRAS, 371, 356

\bibitem[de Bruyn et al. (1996)]{db96}
        de Bruyn, A.G., O'Dea, C.P., \& Baum, S.A. 1996, A\&A,
        305, 450

\bibitem[Dodonov et al. 1999]{dod99}
        Dodonov, S. N.,  Pariiskii, Y. N., Goss, W. M.,
	Kopylov, A. I.,  Soboleva, N. S.,  Temirova, A. V.,
	Verkhodlanov, O. V., Zhelenkova O.P., 1999, ARep, 43, 275

\bibitem[Douglas et al (1996)]{txs96}
        Douglas, J. N., Bash, F. N., Bozyan, F. A., Torrence, G. W.; 
	Wolfe, C., 1996, AJ, 111, 1945

\bibitem[Ellison et al. (2007)]{pair}
        Ellison, S. L., Hennawi, J. F., Martin, C. L., Sommer-Larsen, J.,
	2007, MNRAS, 378, 801

\bibitem[Ellison et al. (2001)]{corals1}
	Ellison, S. L., Yan, L., Hook, I., Pettini, M., Wall, J., Shaver, P.,
	2001, A\&A, 379, 393

\bibitem[Ellison et al. (2002)]{corals2}
	Ellison, S. L., Yan, L., Hook, I., Pettini, M., Wall, J., Shaver, P.,
	2002, A\&A, 383, 91

\bibitem[Erb et al. (2006)]{erb06}
        Erb, D. K., Shapley, A. E., Pettini, M., Steidel, C. C., 
	Reddy, N. A., Adelberger, K. L.,  2006, ApJ, 644, 813 

\bibitem[Ge \& Bechtold (1997)]{gb97}
        Ge, J., Bechtold, J., 1997, ApJ, 477, L73

\bibitem[Ge et al. 2001]{gbk01}
        Ge, J., Bechtold, J., Kulkarni, V., 2001, ApJ, 547, L1

\bibitem[Gehrels, N. (1986)]{geh86}
        Gehrels, N., 1986, ApJ, 303, 336

\bibitem[Haehnelt, Steinmetz and Rauch (1998)]{hsr98}
	Haehnelt, M. G., Steinmetz, M., \& Rauch, M.
	1998, ApJ, 495, 647

\bibitem[Jorgenson et al. (2006)]{jorg06}
        Jorgenson, R., Wolfe, A. M., Prochaska, J. X., Lu, L., Howk, J. C.,
	Cooke, J., Gawiser, E., Gelino, D., 2006, ApJ, 646, 730

\bibitem[Kanekar \& Briggs (2004)]{kb04}
         Kanekar, N., Briggs, F. H., 2004, NewAR, 48, 1259

\bibitem[]{kc01}
        Kanekar, N., \& Chengalur, J.N. 
	2001, A\&A, 369, 42

\bibitem[]{kc03}
        Kanekar, N., \& Chengalur, J.N. 2003, A\&A, 399, 857

\bibitem[Kanekar et al. (2007)]{kan07}
        Kanekar, N., Chengalur, J. N., Lane, W. M., 2007, MNRAS, 
	375, 1528

\bibitem[Kanekar et al. (2006)]{kan06}
        Kanekar, N., Subrahmanyan, R., Ellison, S. L., Lane, W. M., 
	Chengalur, J. N., 2006, MNRAS, 370, L46

\bibitem[Kulkarni \& Heiles (1988)]{kh88}
        Kulkarni, S. R., \& Heiles, C., 1988, Galactic and 
	extragalactic radio astronomy (2nd edition)
	Berlin and New York, Springer-Verlag, pages 95-153.

\bibitem[Lane et al. (1998)]{wl98}
         Lane, W., Smette, A., Briggs, F., Rao, S., Turnshek, D., 
	 Meylan, G. 1998, AJ, 116, 26

\bibitem[Ledoux et al. (2003)]{led03}
        Ledoux, C., Petitjean, P., Srianand, R., 2003, MNRAS,
	346, 209

\bibitem[Liszt (2001)]{hl01}
        Liszt, H. S., 2001, A\&A, 371, 698

\bibitem[Liu et al. 2008]{liu08}
        Liu, X., Shapley, A. E., Coil, A. L., Brinchmann, J., Ma, C.-P.,
	2008, ApJ, accepted	

\bibitem[Lodders (2003)]{lod03}
        Lodders, K., 2003, ApJ, 591, 1220

\bibitem[Lopez et al. (2002)]{lop02}
        Lopez, S., Reimers, D., D'Odorico, S., Prochaska, J. X.,
	2002, A\&A, 385, 778

\bibitem[Lopez et al. (2005)]{seb05}
        Lopez, S., Reimers, D., Gregg, M. D., Wisotzki, L., Wucknitz, O., 
	Guzman, A., 2005, ApJ, 626, 767 

\bibitem[Maller et al. (2001)]{mal01}
        Maller, A. H., Prochaska, J. X., Somerville, R. S., Primack, J. R.,
	2001, MNRAS, 326, 1475

\bibitem[M\o ller, Fynbo \& Fall (2004)]{mff04}
        M\o ller, P., Fynbo, J. P. U., \& Fall, S. M., 2004,
	A\&A, 422, L33
 
\bibitem[Noterdaeme et al. (2007)]{not07a}
        Noterdaeme, P., Ledoux, C., Petitjean, P.,  
	Le Petit, F., Srianand, R., Smette, A.,  2007a, A\&A, 474, 393
 
\bibitem[Noterdaeme et al. (2007)]{not07b}
        Noterdaeme, P., Petitjean, P., Srianand, R., Ledoux, C., 
	Le Petit, F.,  2007b, A\&A, 469, 425
  
\bibitem[Noterdaeme et al. (2008)]{not08}
        Noterdaeme, P., Ledoux, C., Petitjean, P.,  
	Srianand, R.,   2008, A\&A, 481, 327

\bibitem[Pettini et al 1997]{pet97}
        Pettini, M.,  Smith, L.J., King, D.L., \& Hunstead, R.W. 
        1997, ApJ, 486, 665

\bibitem[Prochaska and Wolfe (1997)]{pw97}
	Prochaska, J. X., \& Wolfe, A. M.
	1997, ApJ, 487, 73 

\bibitem[Prochaska et al (2002)]{gb1759}
        Prochaska, J. X., Howk, J. C., O'Meara, J. M., Tytler, D.,
	Wolfe, A. M., Kirkman, D., Lubin, D., Suzuki, N., 2002, ApJ, 571, 693

\bibitem[Prochaska et al (2003)]{jxp03}
        Prochaska, J. X., Gawiser, E., Wolfe, A. M., Castro, S., 
	Djorgovski, S. G., 2003, ApJ, 595, L9

\bibitem[Prochaska, Herbert-Fort \& Wolfe (2005)]{phw05}
        Prochaska, J. X., Herbert-Fort, S., \& Wolfe, A. M., 
	2005, ApJ, 635, 123

\bibitem[Prochaska et al. (2008)]{jxp08}
        Prochaska, J. X., Chen, H.-W., Wolfe, A. M., 
	Dessauges-Zavadsky, M., Bloom, J. S., 2008, ApJ, 672, 59

\bibitem[Prochaska, Hennawi \& Herbert-Fort (2008)]{phhf08}
        Prochaska, J. X., Hennawi, J. F., \& Herbert-Fort, S.,
	2008, ApJ, 675, 1002

\bibitem[Russell, Ellison \& Benn (2006)]{reb06}
        Russell, D., Ellison, S. L., Benn, C. R., 2006, MNRAS, 367, 412

\bibitem[Savaglio et al. (2005)]{sav05}
        Savaglio, S., et al.,  2005, ApJ, 635, 260 

\bibitem[Srianand et al. (2005)]{sri05}
        Srianand, R., Petitjean, P., Ledoux, C., Ferland, G., 
	Shaw, G., 2005, MNRAS, 362, 549

\bibitem[Tripp et al. (2005)]{tripp05}
        Tripp, T. M., Jenkins, E. B., Bowen, D. V., Prochaska, J. X., 
	Aracil, B., Ganguly, R., 2005, ApJ, 619, 714

\bibitem[Tumlinson et al. (2002)]{tum02}
        Tumlinson, J., et al., 2002, ApJ, 566, 857

\bibitem[Wolfe \& Davis (1979)]{wd79}
        Wolfe, A. M., \& Davis, M. M., 1979, AJ, 84, 699

\bibitem[Wolfe, Briggs  \& Jauncey(1981)]{wb81}
         Wolfe, A. M., Briggs, F. H., Jauncey, D. L., 1981, ApJ, 248, 460

\bibitem[Wolfe et al (1985)]{wol85}
        Wolfe, A. M., Briggs, F. H., Turnshek, D. A., Davis, M. M., 
	Smith, H. E., Cohen, R. D., 1985, ApJ, 294, L67


\bibitem[]{}
        Wolfe, A.M., Prochaska, J.X., \& Gawiser, E., 2003, ApJ, 593, 215

\bibitem[Wolfe et al (2008)]{wol08}
	Wolfe, A. M., Prochaska, J. X., Jorgenson, R. A., Rafelski, M.,
	2008, ApJ, accepted

\bibitem[York et al. (2007)]{bay07}
        York, B. A., Kanekar, N., Ellison, S. L., Pettini, M.,
	2007, MNRAS, 382, 53

\bibitem[Young \& Lo (1997)]{yl97}
        Young, L., Lo, K., 1997, ApJ, 490, 710


\end{thebibliography}
\end{document}